\documentclass[12pt]{article}
\usepackage{dcolumn}
\usepackage{graphicx}
\usepackage{epsfig}
\begin{document}
\newfont{\elevenmib}{cmmib10 scaled\magstep1}%
\renewcommand{\theequation}{\arabic{section}.\arabic{equation}}
\newcommand{\tabtopsp}[1]{\vbox{\vbox to#1{}\vbox to12pt{}}}

\newcommand{\preprint}{
            \begin{flushleft}
   \it {~~}
            \end{flushleft}\vspace{-1.3cm}
            \begin{flushright}\normalsize  \sf
            \end{flushright}}
\newcommand{\Title}[1]{{\baselineskip=26pt \begin{center}
            \Large   \bf #1 \\ \ \\ \end{center}}}
\newcommand{\Author}{\begin{center}\large \bf
      S.\, Pratik\, Khastgir
	    \end{center}}
%\hspace*{2.13cm}%
\hspace*{0.7cm}%
\newcommand{\Address}{\begin{center} \it
            Department of Physics and Meteorology and Centre for
	    Theoretical Studies,\\
            Indian Institute of Technology,\\
            Kharagpur 721302, India 
      \end{center}}
\baselineskip=18pt

\preprint
\thispagestyle{empty}
\bigskip
\bigskip
\Title{Affine Toda field theory from tree unitarity}
\Author

\Address
\vspace{1.5cm}

\begin{abstract}
Elasticity property ( i.e. no-particle creation ) is used 
in the tree level scattering of scalar particles in 1+1 dimensions to 
construct affine Toda field
theory(ATFT) associated with the root systems of groups 
$a_2^{(2)}$ and $c_2^{(1)}$. 
A general prescription is given for constructing ATFT
(associated with rank two root systems)
with two self conjugate scalar fields.
It is conjectured that the same method could be used to obtain
the other ATFT associated with higher rank root systems.

\bigskip


\end{abstract}

%%\bigskip
%%\bigskip
%%\bigskip

%%%%%%%%%%%%%%%%%%%%%%%%%%%%%%%%%%%%%%%%%%%%%%%%%%%%%%%%%%%%%%%%%%%%%%
%AUTHOR'S MACROS \& DEFINITIONS
\renewcommand{\theequation}{\arabic{section}.\arabic{equation}}
\def\ts{\thinspace}
\newcommand\et{{\sl et al. }}
\newcommand{\mbold}{\mbox{\boldmath$m$}}
\newcommand{\phibold}{\mbox{\boldmath$\phi$}}
\newcommand{\albold}{\mbox{\boldmath$\alpha$}}
\newcommand{\phibolds}{\mbox{\boldmath$\scriptstyle\phi$}}
\newcommand{\albolds}{\mbox{\boldmath$\scriptstyle\alpha$}}
\newcommand\beq{\begin{equation}}
\newcommand\eeq{\end{equation}}
\newcommand\Bear{\begin{eqnarray}}
\newcommand\Enar{\end{eqnarray}}
\newcommand{\rref}[1]{(\ref{#1})}
\def\hcd#1{\{ #1 \}'{}}
\newcommand\Zam{Zamolodchikov}
\newcommand \ZZ {A. B. \Zam\  and Al. B. \Zam, {\it Ann. Phys.}
{\bf 120} (1979)  253}
\newcommand\CMP{{\it Comm.\ts Math.\ts Phys.\ts}}
\newcommand\IJMP{{\it Int.\ts J.\ts Mod.\ts Phys.\ts}}
\newcommand\NP{{\it Nucl.\ts Phys.\ts}}
\newcommand\PL{{\it Phys.\ts Lett.\ts}}
\newcommand\Zm{Zamolodchikov}
\newcommand\AZm{A.\ts B.\ts \Zm}
\newcommand\dur{H.\ts W.\ts Braden, E.\ts Corrigan, P.\ts E.\ts Dorey 
and R.\ts Sasaki}

%%%%%%%%%%%%%%%%%%%%%%%%%%%%%%%%%%%%%%%%%%%%%%%%%%%%%%%%%%%%

\section{Introduction:}
\setcounter{equation}{0}

 The present note is motivated by the opening section of the paper
``Exact S-matrices'' by Patrick Dorey \cite{Dorey}, which in turn
was inspired by a remark in an article by Goebel on the sine-Gordon
S-matrix \cite{Goebel}. 
 
The aim of this paper is construction of affine Toda field theory (ATFT)
from well known scalar field theory by demanding elasticity property
(i.e. no particle production) in the scattering of particles at tree level.
We would show that the tree-level calculation would suffice for this
purpose. Once the coupling ratios are determined the higher order
elasticity follows. We will see that the three-point couplings
(a ``fusing rule'' for which was proposed in Ref. \cite{Dorey2}) 
play an important role in this.  

In the following we give a very brief description of affine Toda field 
theory.
Affine Toda field theory\footnote{For an excellent review see
Ref. \cite{Co} }\cite{MOPa} is a massive scalar field theory with 
exponential interactions in $1+1$ dimensions described by the Lagrangian,

\begin{equation}
{\cal L}={1\over 2}
\partial{\phibold}\cdot\partial{\phibold}-{m^2\over
\beta^2}\sum_{i=0}^rn_ie^{\beta{\albolds_i\cdot\,\phibolds}}.
\label{ltoda}
\end{equation}
The field $\phibold$ is an $r$-component scalar field, $r$ is the rank of a
compact semi-simple Lie algebra $G$. $\albold_i$;
$i=1,\ldots,r$ are simple roots and $\albold_0$ is 
the affine root of $G$.  The roots are normalized so that long 
roots have length $\sqrt 2$, i.e. $\albold_L^2=2$.
The Kac-Coxeter labels $n_i$ are such that $\sum_{i=0}^rn_i\albold_i=0$, 
with the convention $n_0=1$. The quantity, $\sum_{i=0}^rn_i$, 
is denoted by `$h$' and known as the Coxeter number.
`$m$' is a real parameter setting the mass scale of the theory 
and $\beta$ is a real coupling constant,
which is relevant only in the quantum theory.

ATFT is the best theoretical
laboratory for understanding quantum field theory `beyond
perturbation'. ATFT with real coupling is
one of the best understood field theories at classical and quantum
levels. ATFT is integrable at the classical level
\cite{MOPa,OTa}
due to the presence of an infinite number of conserved quantities. 
Based on the assumption that the infinite set of conserved quantities
be preserved after quantization, only the elastic processes are
allowed and the multi-particle $S$-matrices are factorized into 
a product of two particle elastic $S$-matrices \cite{ZamZam}.
In ATFT, it is well-known that these 
conserved quantities are related with 
the Cartan matrix of the associated finite Lie algebra.
Higher-spin quantum conserved currents are discussed in
Ref. \cite{DGZa}. 
Exact quantum $S$-matrices for all 
simply laced ATFT were evaluated in Refs. [9-14]. 
%%\cite{AFZa}--\cite{KM} 
Most of the non-simply laced  ATFT  exact $S$-matrices
 were calculated in Ref. \cite{DGZc} with the 
beautiful idea of floating masses. These $S$-matrices
respect crossing symmetry and bootstrap principle
\cite{ZamZam,BCDSc}.
%%\cite{AFZa,BCDSa,BCDSc,CMa,DDa,KM}.
The exact quantum $S$-matrices for the remaining non-simply laced
theory  were constructed in Ref. \cite{CDS} where 
generalized bootstrap principle was introduced and more insight
to the mechanism was provided. 
The singularity structure of the  
$S$-matrices of simply laced theories, which in some   
cases contain poles up to 12-th order \cite{BCDSc},
is  beautifully explained in terms of the
singularities of the corresponding Feynman diagrams
\cite{BCDSe}, so called Landau singularities.
Finally Affine Toda field theory is one place where one can see explicitly 
the recently popular strong-weak coupling duality. It is known that
exact Toda $S$-matrices for simply laced systems are invariant under 
this duality.  

Next section presents the results obtained in the opening section
of Ref. \cite{Dorey}. Section 3, solves the exercise suggested at
end of the opening section of the Ref. \cite{Dorey} to obtain
the $a_2^{(2)}$ ATFT or the Bullough-Dodd theory. Section 4, works
with two scalar fields of different masses(one has a mass $\sqrt 2$
times the other) and an interaction between them. This problem leads 
to an ATFT associated with $c_2^{(1)}$ root system, which is non-simply 
laced. Section 5 will deal with a general approach towards theories with
two scalar particles which are self conjugate.
In this section the allowed values of mass ratio and 3-point couplings
would be obtained and the final theory would come out to be an
ATFT associated with a rank two root system.
Section 6 is reserved for conclusions and a conjecture.

\section{ sinh-Gordon or $a_1^{(1)}$ theory }
\setcounter{equation}{0}

This section is shamelessly lifted from the ``Introduction'' of the
paper \cite{Dorey}. Starting from scalar $\phi^4$ theory in 1+1 dimensions
the simplest possible ATFT i.e. sinh-Gordon or $a_1^{(1)}$ is
obtained.

We begin with1+1 D scalar $\phi^4$ Lagrangian, 

\begin{equation} {\cal L}=\frac{1}{2}(\partial
\phi)^{2}-\frac{1}{2}m^{2}\phi^{2}-\frac{\lambda}{4!}\phi^{4}.
\label{phi4}
\end{equation} 

The Feynman rules are:

\begin{center}
${}$\hspace {.2in}\scalebox{.5}
{\includegraphics{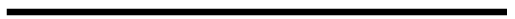}} = $ \displaystyle\frac{i}{p^{2}-m^{2}
+i\epsilon}$
\end{center}

\begin{center}
${}$\hspace {.2in}
\scalebox{.3}{\includegraphics{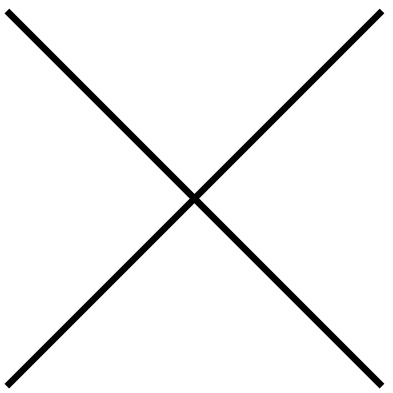} } 
\hspace{.2in}\raisebox{3.5ex}{= $-i\lambda$}
\end{center}
where $p$ is momentum and $m$ is the mass of the particles. 
  We use light-cone coordinates,
 $$(p,\bar{p})=(p^{0}+p^{1},p^{0}-p^{1}).$$ 

\begin{figure}[h]
%%\vspace{-1.7in}
${}$\hspace {.2in}\scalebox{.6}
{\includegraphics{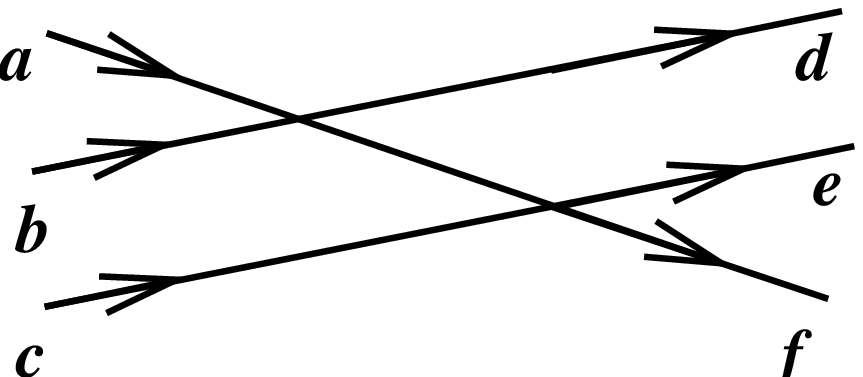}\hspace {2.0in}\includegraphics{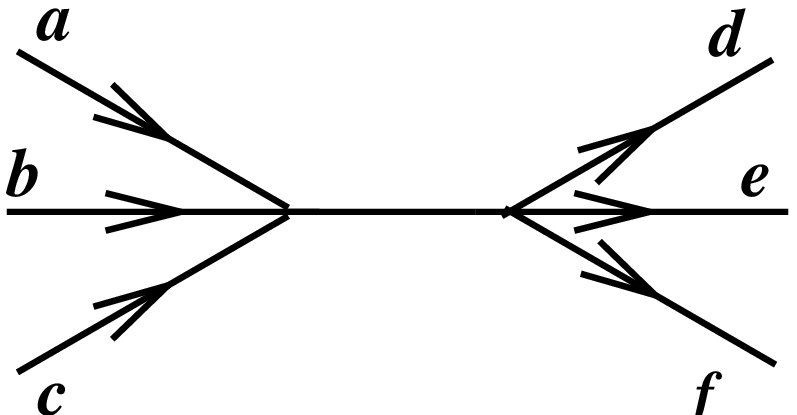}}
\caption{ (a) and (b), 3$\phi\rightarrow 3\phi$ process}
\label{3to3}
\end{figure}

\noindent Using the mass-shell condition 
$p\bar{p}=m^{2}$, \emph{in} and \emph{out}
momenta are written as
$$(p_a,\bar{p_a})=(ma,ma^{-1}),~~~(p_b,\bar{p_{b}})=(mb,mb^{-1})$$ and
so on, with $a, b,\dots $  real numbers, positive for particle 
traveling forward in time.
One now calculates the connected  $2\phi\rightarrow 4\phi$ production 
amplitude at tree level. For this one looks at the diagrams  of 
$3\phi\rightarrow 3\phi$ processes, with
implicit understanding that one of the \emph{in} momenta will be
crossed to \emph{out} at the end. The \emph{in} particles are labeled as
$a, b, c$ and the \emph{out} particles as $d, e, f$. In terms of these
variables crossing from $3\phi\rightarrow 3\phi$ to   
$2\phi\rightarrow 4\phi$ amounts
to a continuation from $c$ to $-c$.
For the $3\phi\rightarrow 3\phi$ amplitude at tree level there are just the
following two classes of diagrams (Fig. \ref{3to3}) as shown in 
the Ref. \cite{Dorey},

\bigskip

As one of the  \emph{in} momenta is actually going \emph{out}, thus the
propagator is not on the mass-shell so removal of $i\epsilon$ terms is
allowed. Thus the internal momentum
$p=m(a+b-d,a^{-1}+b^{-1}-d^{-1})$, and the contribution to the
propagator  from above Fig. \ref{3to3} (a) is

\begin{equation}
\frac{i}{p^{2}-m^{2}}=\frac{i}{m^{2}[(a+b-d)(a^{-1}+b^{-1}-d^{-1})-1]}
=\frac{-iabd}{m^{2}(a+b)(a-d)(b-d)}.
\end{equation}
Similarly for the Fig. \ref{3to3} (b) the contribution to the propagator is:

\begin{equation}
\frac{i}{p^{2}-m^{2}}=\frac{iabc}{m^{2}(a+b)(a+c)(b+c)}.
\end{equation}
Taking all the terms in accounts the amplitude of $\emph{in}
\rightarrow \emph{out}$
is:
\begin{equation}
{\langle {out\,}|{\,in}\rangle}_{\rm tree}=
-\frac{i\lambda^{2}}{m^{2}}A_{legs}H(a,b,c,d,e,f),
\end{equation}
where $A_{legs}$ contains all the common factors on external legs,
and
\begin{equation}
H(a,b,c,d,e,f)=\left[\sum_{{cycl\{a,b,c\}}\atop{cycl\{d,e,f\}}}
\frac{-abd}{(a+b)(a-d)(b-d)}\right]+\frac{abc}{(a+b)(b+c)(c+a)}.
\end{equation}
Using $a+b+c=d+e+f$ and
$a^{-1}+b^{-1}+c^{-1}=d^{-1}+e^{-1}+f^{-1}$. i.e.
the  conservation of left- and
right-light-cone momenta respectively, one finds 
$H(a,b,c,d,e,f)=-1$.
As above argument does not contain the sign of any momenta, it 
holds for $-c$ also.

So we find in $1+1$ D $\lambda \phi^{4}$ theory, the amplitude of
$2\phi\rightarrow 4\phi$ is constant at tree level.

\noindent By adding
a term $\displaystyle -\frac{\lambda^{2}}{6!m^{2}}\phi^{6}$ to the original
Lagrangian (\ref{phi4}) one can make the $2\phi \rightarrow 4\phi$
amplitude to vanish. Defining  $\beta^2=\lambda/m^2$, the new
Lagrangian up to $\phi^6$ order is,

\begin{equation} {\cal L}=\frac{1}{2}(\partial
\phi)^{2}-\frac{m^2}{\beta^2}\left[\frac{1}{2}\beta^2\phi^{2}+
\frac{1}{4!}\beta^4\phi^{4}+\frac{1}{6!}\beta^6\phi^{6}\right].
\label{phi6}
\end{equation}

Now one calculates $2\phi\rightarrow 6\phi$ tree
level amplitude with this  
$\displaystyle -\frac{\lambda^{2}}{6!m^{2}}\phi^{6}$
term added to the Lagrangian (\ref{phi4}) and finds it to be a
constant, which can be canceled by a judiciously chosen $\phi^8$
term and so on. At each stage a residual constant piece can be removed
by a (uniquely determined) higher-order interaction.
 After adding infinitely many terms in this way assuring no particle
production at tree level one finds,
$${\cal L}=\frac{1}{2}(\partial\phi)^{2}-\frac{m^{2}}{\beta^{2}}
[\frac{1}{2!}\beta^{2}\phi^{2}+
\frac{1}{4!}\beta^{4}\phi^{4}+\frac{1}{6!}\beta^{6}\phi^{6}
+\frac{1}{8!}\beta^{8}\phi^{8}\dots]$$
\begin{equation}
=\frac{1}{2}(\partial
\phi)^{2}-\frac{m^{2}}{\beta^{2}}\left[\cosh(\beta \phi)-1\right]=
{\cal L}_{\mbox{\boldmath{$\scriptstyle a_1^{(1)}$}}}.
\end{equation}
The above Lagrangian, the simplest ATFT, is
sinh-Gordon or $a_1^{(1)}$ Lagrangian
and is well studied in the literature.
Araf'eva and Korepin showed in Ref. \cite{Arkor} that the elasticity
is maintained at one loop level for the above Lagrangian.
The well known sine-Gordon Lagrangian could be obtained 
by sending the coupling $\beta$ to imaginary.

\section{Bullough-Dodd model or $a_2^{(2)}$ theory}
\setcounter{equation}{0}

What would happen if we played the same game with a $\phi^3$ theory?
This was suggested as an exercise in \cite{Dorey}. The
solution follows here in detail.
The Lagrangian we begin with has the following form,

\begin{equation} 
{\cal L}=\frac{1}{2}(\partial
\phi)^{2}-\frac{1}{2}m^{2}\phi^{2}-\frac{\eta}{3!}\phi^{3}.
\label{phi3}
\end{equation} 

Again the Feynman rules are:

\begin{center}
${}$\hspace {.2in}\scalebox{.5}
{\includegraphics{propa.eps}} = $\displaystyle 
\frac{i}{p^{2}-m^{2}+i\epsilon}$
\end{center}

\begin{center}
${}$\hspace {.2in}
\scalebox{.3}{\includegraphics{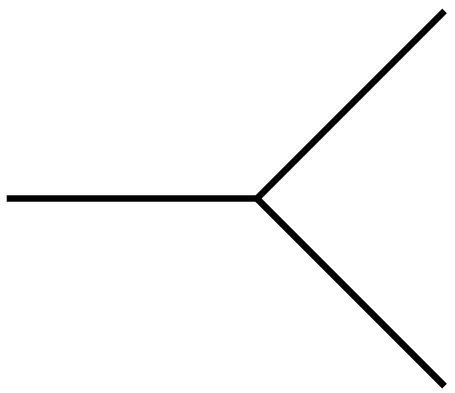} } 
\hspace{.2in}\raisebox{3.5ex}{= $-i\eta$}
\end{center}

\begin{figure}[h]
\begin{center}
${}$\hspace {.2in}\scalebox{.5}
{\includegraphics{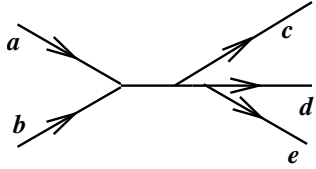}}
\caption{2$\phi\rightarrow3\phi$ process with three $\phi^3$ vertices}
\label{2phto3ph}
\end{center}
\end{figure}

First we consider a $2\phi\rightarrow 3\phi$ process for which we have the
following tree level diagram (Fig. \ref{2phto3ph}).
For making our calculations easier we take both $\emph{in}$ momenta
(($a$, $a^{-1}$) and ($b$, $b^{-1}$))
equal to (1,1) and one of the $\emph{out}$ momenta, ($e, e^{-1}$), equal to
($1+\delta, (1+\delta)^{-1}$), $\delta$ need not be small. 

Now the conservation of left- and right-light-cone momenta would give,
\begin{equation}
c+d=1-\delta,\quad c^{-1}+d^{-1}={{1+2\delta}\over{1+\delta}},
\quad cd={{1-\delta^2}\over{1+2\delta}}, \quad cd^{-1}+c^{-1}d=
-{{1+\delta+2\delta^2}\over{1+\delta}},
\label{consv}
\end{equation}
where ($c, c^{-1}$) and ($d, d^{-1}$) are momenta of other two
outgoing particles.
There are altogether fifteen diagrams of the above type (details of
their individual contributions are given in appendix A1). Summing all
the diagrams using above relations, (\ref{consv}), we obtain,

\begin{equation}
{{i\eta^3}\over{m^4}}\left[- ~{3\over2}{{(1+\delta)}\over {\delta^2}} -1
 + {3\over2}{{(4+\delta)}\over {(1+\delta+\delta^2)}}
+ {9\over2}{{\delta}\over {(1+\delta+\delta^2)^2}}\right]. 
\label{33ampl}
\end{equation}

\begin{figure}[h]
%%\vspace{-1.7in}
\begin{center}
${}$\hspace {.2in}\scalebox{.6}
{\includegraphics{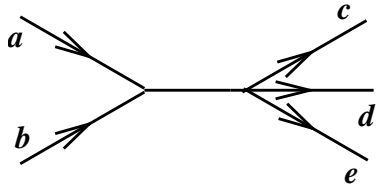}}
\caption{2$\phi\rightarrow3\phi$ process with a $\phi^3$ vertex and a
$\phi^4$ vertex}
\label{2pto3p}
\end{center}
\end{figure}

For stopping the particle production at tree level we add a counter
term $\displaystyle -{\lambda\over 4!}\phi^4$ to the 
Lagrangian (\ref{phi3}). This
would produce a new Feynman rule,

\begin{center}
${}$\hspace {.2in}
\scalebox{.3}{\includegraphics{vertex4.eps} } 
\hspace{.2in}\raisebox{3.5ex}{= $-i\lambda$,}
\end{center}
giving the following class of new diagrams (Fig. \ref{2pto3p})
for the tree level $2\phi\rightarrow 3\phi$ process.

There are total of ten diagrams of the above type and we sum them
again using the relations (\ref{consv}). The total contribution is
(for individual details and contributions of the diagrams see appendix A2),

\begin{equation}
{{i\eta\lambda}\over{m^2}}\left[{1\over 2}{{(1+\delta)}\over{\delta^2}} 
+ 2 - {1\over 2}{{(4+\delta)}\over {(1+\delta+\delta^2)}}
- {3\over2}{{\delta}\over {(1+\delta+\delta^2)^2}}\right]. 
\label{34ampl}
\end{equation}

Adding (\ref{34ampl}) and  (\ref{33ampl}) we obtain total tree level
contribution of the $2\phi\rightarrow 3\phi$ process with $\phi^3$ and $\phi^4$
terms in the Lagrangian as,

\begin{equation}
\scalebox{.4}
{\includegraphics{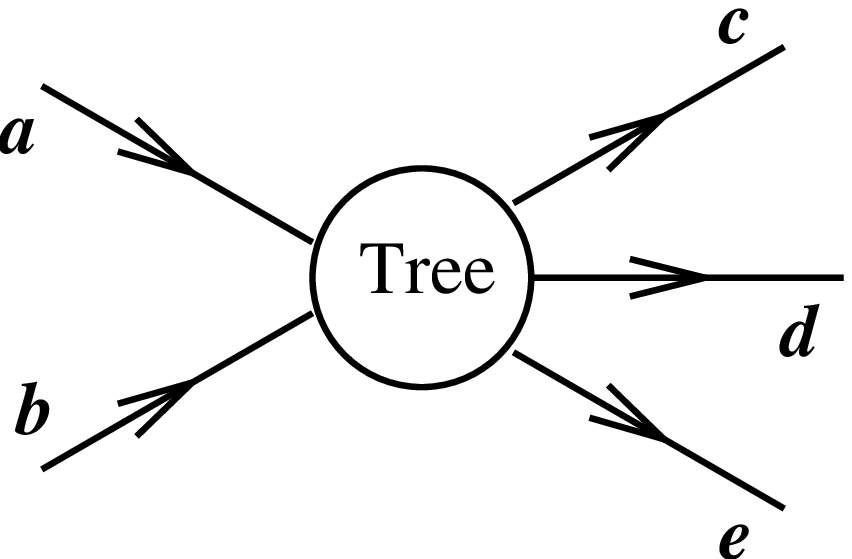}}\hspace{.2in}
\raisebox{5.8ex}{$=({{i2\eta\lambda}\over{m^2}}
-{{i\eta^3}\over{m^4}})
+({{i\eta\lambda}\over{m^2}}-{{i3\eta^3}\over{m^4}})
\left[{1\over 2}{{(1+\delta)}\over{\delta^2}} 
- {1\over 2}{{(4+\delta)}\over {(1+\delta+\delta^2)}}
- {3\over2}{{\delta}\over {(1+\delta+\delta^2)^2}}\right].$}
\end{equation}

Now if one chooses $\lambda={\displaystyle{{3\eta^2}\over m^2}}$, 
one gets rid of all
the terms involving parameter $\delta$ and the total contribution
becomes a constant equal to $\displaystyle{{i5\eta^3}\over m^4}$.
This constant contribution could be killed if another counter term,
$-\displaystyle{{5\eta^3}\over{5! m^4}}\phi^5$, is added to the
Lagrangian (\ref{phi3}). After adding $\phi^4$ and $\phi^5$ terms, 
the new Lagrangian looks like,

\begin{equation} 
{\cal L}=\frac{1}{2}(\partial
\phi)^{2}-\frac{1}{2}m^{2}\phi^{2}-\frac{\eta}{3!}\phi^{3}
-\frac{3\eta^2}{4!m^2}\phi^{4}-\frac{5\eta^3}{5!m^4}\phi^{5} \cdots
\label{phi3m}
\end{equation} 

The above Lagrangian, (\ref{phi3m}) would produce vanishing result
for the tree level $2\phi\rightarrow 3\phi$ process. 
To fix the dots in the above
Lagrangian one should look for $2\phi\rightarrow 4\phi$ tree level process.
Next one demands vanishing contribution for this $2\phi\rightarrow
4\phi$ tree level process to fix the $\phi^6$ term and one 
can proceed so on order by order.

Setting $\beta=\displaystyle{\eta\over m^2}$ we observe that 
 the Lagrangian, (\ref{phi3m}), contains first four terms of the
 following Lagrangian after a power series expansion,

\begin{equation} 
{\cal L}_{\mbox{\boldmath{$\scriptstyle a_2^{(2)}$}}}=\frac{1}{2}(\partial
\phi)^{2}-\frac{m^2}{6\beta^2}\left
    [~e^{2\beta\phi}+2e^{-\beta\phi}-3~\right ].
\label{buldod}
\end{equation} 

The above, (\ref{buldod}) is another well studied Lagrangian known as 
the Bullough-Dodd model
or the $a_2^{(2)}$ ATFT in the literature.
We mention in passing that if one starts with a theory with both $\phi^3$
and $\phi^4$ terms absent in the Lagrangian then no particle
production at tree level would lead to
a free theory that is having all higher couplings vanishing.

\section{$c_2^{(1)}$ theory}
\setcounter{equation}{0}

In this section we consider two interacting self conjugate scalar
fields. The starting Lagrangian in this case is chosen to be

\begin{equation}
{\cal L}=\frac{1}{2}(\partial
\phi_1)^{2}+\frac{1}{2}(\partial
\phi_2)^{2}-m^{2}\phi_1^{2}-\frac{1}{2}m^{2}\phi_2^{2}+
\frac{\xi}{2!}\phi_1\phi_2^{2}.
\label{c21}
\end{equation}
Notice that the particle $\phi_1$ is ${\sqrt 2}$ times heavier than particle
$\phi_2$ and there is only one interaction term, viz. $\phi_1\phi_2^{2}$.

The Feynman rules are the following,
 
\begin{center}
${}$\hspace {.2in}$\phi_1$ propagator~:~~
\scalebox{.4}{\includegraphics{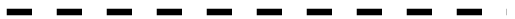}} = 
$\displaystyle \frac{i}{p^{2}-2m^{2}+i\epsilon}$
\end{center}

\begin{center}
${}$\hspace {.1in} $\phi_2$ propagator~:~~
\scalebox{.4}{\includegraphics{propa.eps}}
= $ \displaystyle\frac{i}{p^{2}-m^{2}+i\epsilon}$
\end{center}

\begin{center}
${}$\hspace {.2in}
\scalebox{.3}{\includegraphics{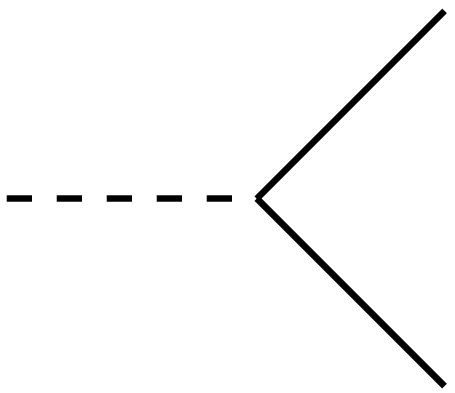} } 
\hspace{.2in}\raisebox{3.1ex}{= $i\xi$}
\end{center}

\begin{figure}[h]
\begin{center}
${}$\hspace {.2in}
\scalebox{.5}{\includegraphics{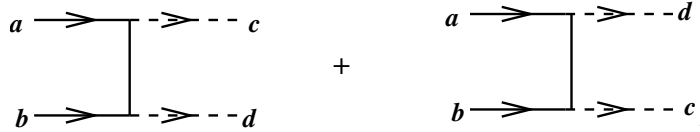}}. 
\end{center}
\caption{$2\phi_2\rightarrow 2\phi_1$ process}
\label{2p2to2p1}
\end{figure}

First we consider the process $2\phi_2\rightarrow 2\phi_1$. At tree
level, there are following two diagrams (Fig. \ref{2p2to2p1}),

The $\emph{in}$ particles have momenta ($a,a^{-1}$) and
($b,b^{-1}$) whereas the $\emph{out}$ ones have momenta 
($\sqrt{2}\,c,\sqrt{2}\,c^{-1}$) and ($\sqrt{2}\,d,\sqrt{2}\,d^{-1}$)
without loss of generality since the  $\emph{out}$ particles are
$\sqrt 2$ times heavy as the $\emph{in}$ ones. The conservation of
left-and right-light-cone momenta gives the following
relations,
\begin{equation}
c+d={{a+b}\over{\sqrt{2}}},\quad
c^{-1}+d^{-1}={{a^{-1}+b^{-1}}\over\sqrt{2}},
\quad cd=ab, \quad cd^{-1}+c^{-1}d +1={1\over 2}(ab^{-1}+a^{-1}b).
\label{consv21}
\end{equation}
 
Individual contributions of the above two diagrams are,
\begin{equation}
{{-i\xi^2}\over {m^2[2-{\sqrt 2}(ac^{-1}+a^{-1}c)]+i\epsilon}}+
{{-i\xi^2}\over {m^2[2-{\sqrt 2}(ad^{-1}+a^{-1}d)]+i\epsilon}}.
\end{equation}
Summing the above two expressions using relations (\ref{consv21})
and then taking the limit $\epsilon\rightarrow 0$ we obtain
a constant equal to $\displaystyle\frac{i\xi^2}{m^2}$. This constant
contribution will be killed if we add a term,
 $\displaystyle-~{1\over {2!2!}}\frac{\xi^2}{m^2}\phi_1^2\phi_2^2$,
 to the Lagrangian (\ref{c21}). This new term gives an additional
Feynman rule,

\begin{center}
${}$\hspace {.2in}
\scalebox{.3}{\includegraphics{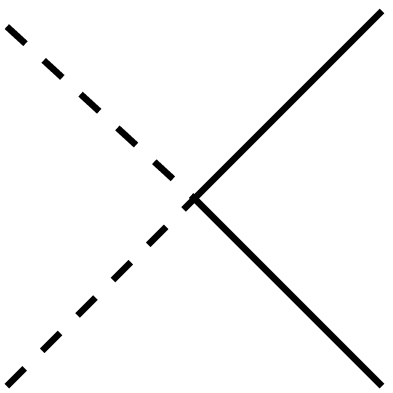} } 
\hspace{.2in}\raisebox{3.5ex}{= $\displaystyle-\frac{i\xi^2}{m^2}$.}
\end{center}

\begin{figure}[h]
\begin{center}
${}$\hspace {.2in}
\scalebox{.5}{\includegraphics{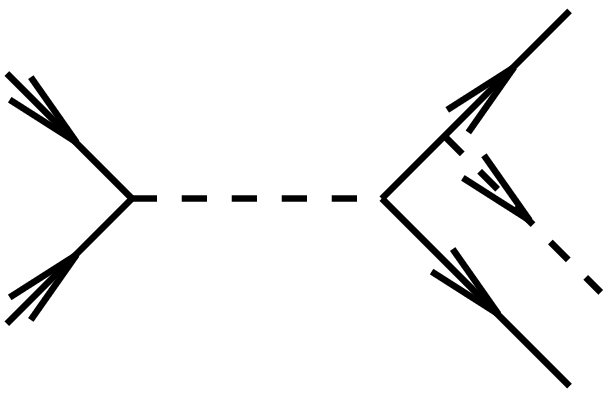} }
\hspace{.3in}\raisebox{5.0ex}{+}\hspace{.3in}
\scalebox{.5}{\includegraphics{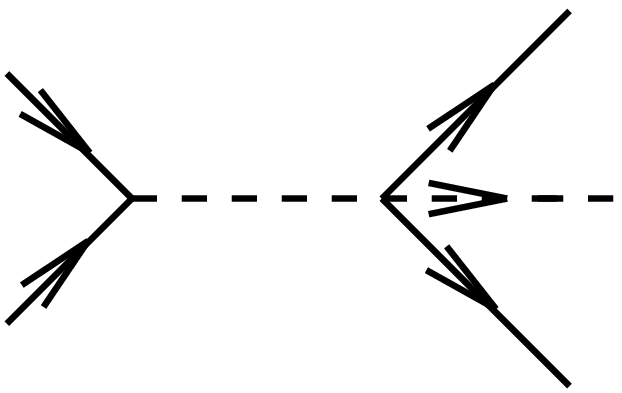} }
\end{center}
\caption{$2\phi_2\rightarrow 2\phi_2+\phi_1$ process (a) with 
3 $\phi_1\phi_2^2$ vertices and (b) with one $\phi_1\phi_2^2$ vertex
and one $\phi_1^2\phi_2^2$ vertex.}
\label{2p2to2p2p1}
\end{figure}

Next we look at the process $2\phi_2\rightarrow 2\phi_2+\phi_1$. Again for
simplifying our calculations we take $\emph{in}$ momenta as (1,1)
and outgoing particle $\phi_1$ momentum as ($\sqrt{2}e, 
\sqrt{2}e^{-1}$). We have the following two types of
diagrams (Fig. \ref{2p2to2p2p1}), one having 
three $\phi_1\phi_2^2$ vertices and the other
containing one $\phi_1\phi_2^2$ vertex and one $\phi_1^2\phi_2^2$
vertex. 

\begin{figure}[h]
\begin{center}
${}$\hspace {.2in}
\scalebox{.5}{\includegraphics{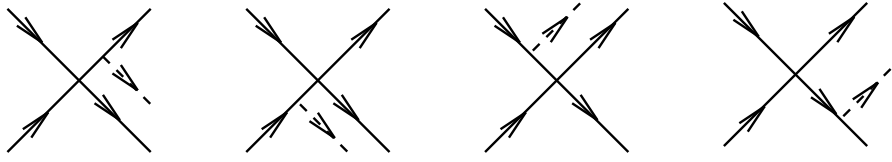} }. 
\end{center}
\caption{$2\phi_2\rightarrow 2\phi_2+\phi_1$ with a $\phi_2^4$ vertex
  and a $\phi_1\phi_2^2$ vertex}
\label{4more}
\end{figure}

\begin{figure}[h]
\begin{center}
${}$\hspace {.2in}
\scalebox{.5}{\includegraphics{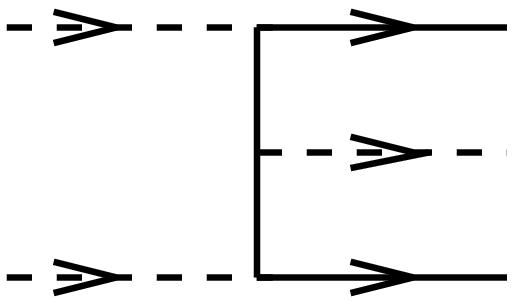} }
\hspace{.3in}\raisebox{3.2ex}{+}
\hspace{.3in}\raisebox{-2.0ex}{\scalebox{.5}{\includegraphics{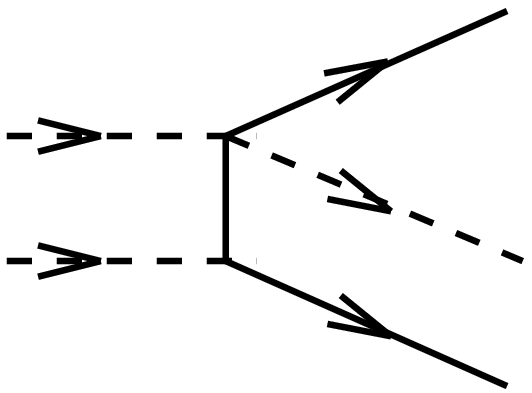} }}
\end{center}
\caption{a) and b) $2\phi_1\rightarrow 2\phi_2+\phi_1$ process}
\label{2more}
\end{figure}

We have 12 diagrams of the former type and 6 diagrams of the
latter type (details of which are presented in the appendix B1).
Summing all 18 diagrams one obtains (using momentum conservation
relations of course),
\begin{equation}
-\frac{i\xi^3}{m^4}
\frac{e\,(8e-3{\sqrt 2}\,e^2-3{\sqrt 2})}{(2e-{\sqrt 2}\,e^2-{\sqrt 2})^2}
\label{21cube}
\end{equation}
Now if one adds a counter term
$\displaystyle{-\frac{\zeta}{4!}\phi_2^4}$ to the Lagrangian,
(\ref{c21}) one will have a new vertex of the following type,

\begin{center}
${}$\hspace {.2in}
\scalebox{.3}{\includegraphics{vertex4.eps} } 
\hspace{.2in}\raisebox{3.5ex}{= $-i\zeta$}
\end{center}
This new vertex in turn would add the following 4 more diagrams
(Fig. \ref{4more}) to the above 
process $2\phi_2\rightarrow 2\phi_2+\phi_1$.
Contribution of these four diagrams when added is equal to,
     
\begin{equation}
-\frac{i\xi\zeta}{m^2} +\frac{i\xi\zeta}{m^2}
\frac{e\,(8e-3{\sqrt 2}\,e^2-3{\sqrt 2})}{(2e-{\sqrt 2}\,e^2-{\sqrt 2})^2}.
\label{421cont}
\end{equation}
Adding (\ref{21cube}) and (\ref{421cont}) we have,

\begin{equation}
\scalebox{.5}{\includegraphics{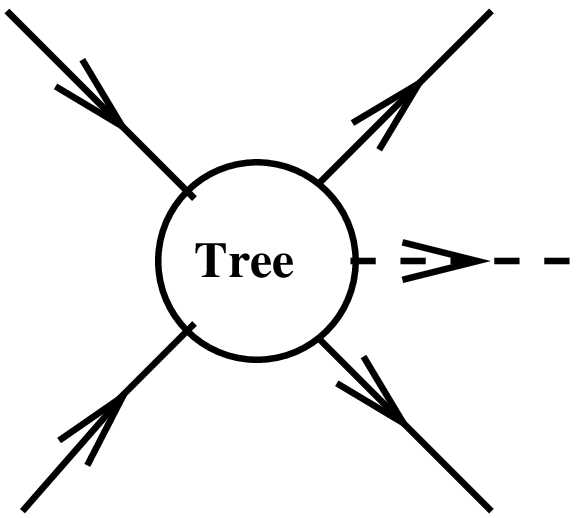} }\hspace{.2in} 
\raisebox{6.8ex}{$\displaystyle = -\frac{i\xi\zeta}{m^2} 
+(\frac{i\xi\zeta}{m^2}-\frac{i\xi^3}{m^4})
\frac{e\,(8e-3{\sqrt 2}\,e^2-3{\sqrt 2})}{(2e-{\sqrt 2}\,e^2-{\sqrt 2})^2}$}
\label{point5}
\end{equation}
The above expression, (\ref{point5}), clearly shows 
if we choose $\displaystyle\zeta={\xi^2\over m^2}$ 
then the above expression becomes independent of the parameter
  $e$ and the total sum becomes $\displaystyle -{i\xi^3\over m^4}$.
This constant contribution will be killed if one adds a new counter
term $\displaystyle\frac{1}{4!}\frac{\xi^3}{m^4}\phi_1\phi_2^4$ to the 
Lagrangian (\ref{c21}). At this stage our Lagrangian reads,

\begin{equation}
{\cal L}=\frac{1}{2}(\partial
\phi_1)^{2}+\frac{1}{2}(\partial
\phi_2)^{2}-m^{2}\phi_1^{2}-\frac{1}{2}m^{2}\phi_2^{2}+
\frac{\xi}{2!}\phi_1\phi_2^{2}
-{1\over {2!2!}}\frac{\xi^2}{m^2}\phi_1^2\phi_2^2
-\frac{1}{4!}\frac{\xi^2}{m^2}\phi_2^4
+\frac{1}{4!}\frac{\xi^3}{m^4}\phi_1\phi_2^4.
\label{c21m}
\end{equation}
To fix the remaining quartic and quintic interactions we concentrate
on $2\phi_1\rightarrow 2\phi_2+\phi_1$ process which possesses following two
classes of diagrams (Fig. \ref{2more}), viz. one with three
 $\phi_1\phi_2^2$ vertices and the other
containing one $\phi_1\phi_2^2$ vertex and one $\phi_1^2\phi_2^2$
vertex like before. 

This case we choose $\emph{in}$ momenta for both the particles
as $({\sqrt 2},{\sqrt 2})$ and outgoing particle $\phi_1$ momentum
is designated by $({\sqrt 2}\,e,{\sqrt 2}\,e^{-1})$. There are 6 diagrams
of each kind (see appendix B2 for details). Summing all 12 diagrams
we obtain,

\begin{equation}
\scalebox{.5}{\includegraphics{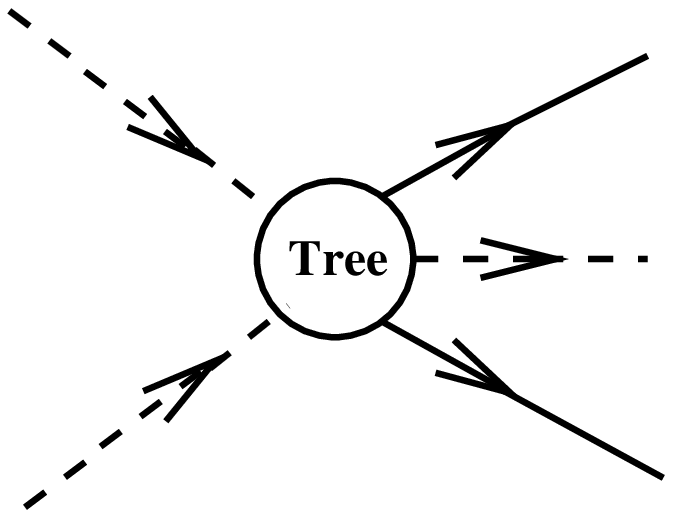} }
\hspace{.3in}\raisebox{6.2ex}{$\displaystyle 
=-\frac{i\xi^3}{m^4}+\frac{i\xi^3}{2m^4}\frac{e}{(e-1)^2}$}.
\label{tree22to221}
\end{equation}
From the above expression (\ref{tree22to221}) it is easy to fix 
the quartic $\phi_1^4$ counter term
so that the parameter $e$ dependent term is killed. For this
we add a term $\displaystyle -\frac{\gamma}{4!}\phi_1^4$ to the
Lagrangian (\ref{c21m}) to obtain the following new diagram for
the above process.
 
\begin{equation}
\scalebox{.5}{\includegraphics{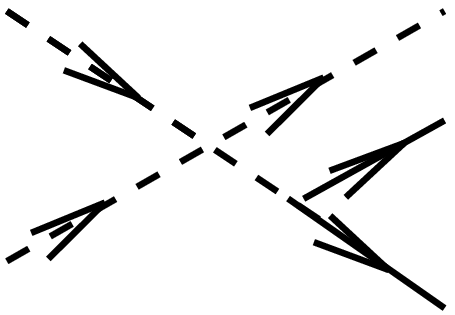} }
\hspace{.3in}\raisebox{4.2ex}{$\displaystyle 
=-\frac{i\xi\gamma}{4m^2}\frac{e}{(e-1)^2}$~.}
\label{phi1422to221}
\end{equation}
Now it is very clear from the above expressions (\ref{tree22to221}) 
and (\ref{phi1422to221}), if we choose $\displaystyle
\gamma=\frac{2\xi^2}{m^2}$ all $e$ dependent terms would cancel
from the tree level $2\phi_1\rightarrow 2\phi_2+\phi_1$ process and the result
would be a constant equal to $\displaystyle -\frac{i\xi^3}{m^4}$.
This constant contribution is canceled by a vertex of the
following type,

\begin{equation}
\scalebox{.5}{\includegraphics{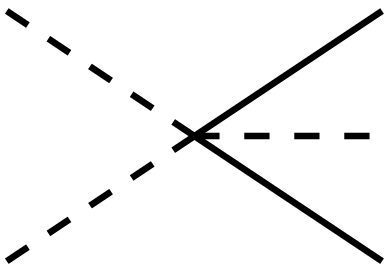} }
\hspace{.3in}\raisebox{3.2ex}{$\displaystyle 
=\frac{i\xi^3}{m^4}~.$}
\end{equation}
The above vertex corresponds to adding a 
$\displaystyle \frac{1}{3!2!}\frac{\xi^3}{m^4}\phi_1^3\phi_2^2$ term to the
Lagrangian (\ref{c21m}). The final Lagrangian with all 5-point
interaction vertices becomes,

\begin{eqnarray}
{\cal L} & = & \frac{1}{2}(\partial
\phi_1)^{2}+\frac{1}{2}(\partial
\phi_2)^{2}-m^{2}\phi_1^{2}-\frac{1}{2}m^{2}\phi_2^{2}+
\frac{\xi}{2!}\phi_1\phi_2^{2}
-\frac{2}{4!}\frac{\xi^2}{m^2}\phi_1^4 \nonumber\\
& & -{1\over {2!2!}}\frac{\xi^2}{m^2}\phi_1^2\phi_2^2
-\frac{1}{4!}\frac{\xi^2}{m^2}\phi_2^4
+\frac{1}{4!}\frac{\xi^3}{m^4}\phi_1\phi_2^4
+\frac{1}{3!2!}\frac{\xi^3}{m^4}\phi_1^3\phi_2^2~.
\label{c21f}
\end{eqnarray}

The above Lagrangian, (\ref{c21f}) contains the first eight terms
(after expansion) of
the following $c_2^{(1)}$ ATFT Lagrangian, 

\begin{equation}
{\cal L}_{\mbox{\boldmath{$\scriptstyle c_2^{(1)}$}}} = \frac{1}{2}\partial
\phibold\cdot\partial
\phibold -\frac{m^2}{2\beta^2}\left[e^{\beta\albolds_0\cdot\phibolds}+
2e^{\beta\albolds_1\cdot\phibolds}+e^{\beta\albolds_2\cdot\phibolds}-4\right],
\end{equation}
where field $\phibold$ has two components i.e. $\phi_1$ and $\phi_2$.
$\albold_0$ is affine and $\albold_1$ and $\albold_2$ are simple 
roots of algebra $c_2^{(1)}$ ($\albold_1=(1,0),~\albold_2=(-1,1),
~\albold_0=(-1,-1)$)
 and $\displaystyle\beta=\frac{\xi}{m^2}$.
This is another integrable model which is well studied \cite{BCDSc}.
Again all the higher n-point couplings $(n>5)$ could be fixed by studying the
various other tree level processes.

\section{Other theories with two self conjugate scalar fields}
\setcounter{equation}{0}

In this section we give a general method for constructing
various other integrable theories associated with rank two root systems.
We have seen in the previous section that the sole three point
interaction decides the fate of other terms if one maintains
elasticity property order by order at tree level.
One can verify that elasticity is maintained if one goes to loop diagrams. 
Here we start with the most general Lagrangian with two
self conjugate scalar fields with all possible three-point
interactions,
\begin{eqnarray}
{\cal L} & = & \frac{1}{2}(\partial
\phi_1)^{2}+\frac{1}{2}(\partial
\phi_2)^{2}-\frac{q}{2}m^{2}\phi_1^{2}-\frac{1}{2}m^{2}\phi_2^{2}\nonumber\\
& &+\frac{\xi}{2!}\phi_1\phi_2^{2}
-\frac{r\xi}{3!}\phi_1^{3}
-\frac{s\xi}{3!}\phi_2^{3}
-\frac{t\xi}{2!}\phi_1^2\phi_2.
\label{genl}
\end{eqnarray}
Note that we have fixed the strength of one mass term and 
only one of the three-point interactions. The other 
mass and couplings have strengths relative to
these. Our objective is to determine these relative strengths
( i.e. $q$, $r$, $s$ and $t$) for an integrable theory.
 Feynman rules are given by,

\begin{center}
${}$\hspace {.2in}$\phi_1$ propagator~:~~
\scalebox{.4}{\includegraphics{propa2.eps}} = 
$\displaystyle \frac{i}{p^{2}-q\,m^{2}+i\epsilon}$
\end{center}

\begin{center}
${}$\hspace {.1in} $\phi_2$ propagator~:~~
\scalebox{.4}{\includegraphics{propa.eps}} = 
$\displaystyle\frac{i}{p^{2}-m^{2}+i\epsilon}$
\end{center}

\begin{center}
${}$\hspace {.2in}
\scalebox{.3}{\includegraphics{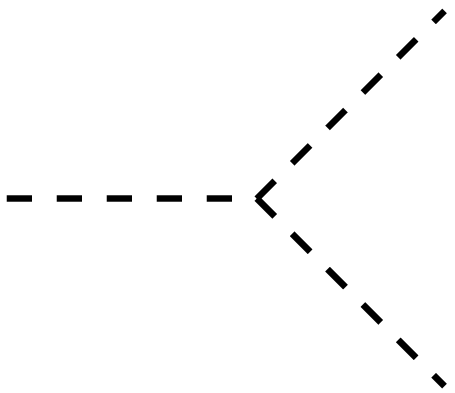} } 
\raisebox{2.8ex}{= $-ir\xi,$}\hspace{.3in}
\scalebox{.3}{\includegraphics{122ver.eps} } 
\raisebox{2.8ex}{= $i\xi,$}\hspace{.3in}
\scalebox{.3}{\includegraphics{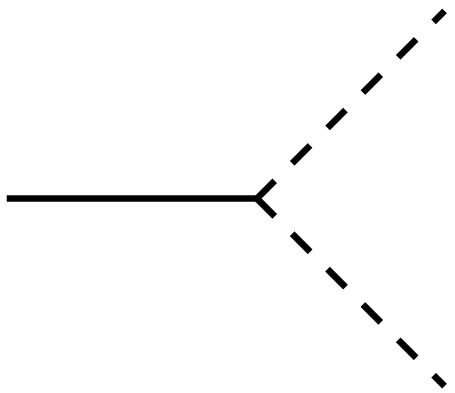} } 
\raisebox{2.8ex}{= $-it\xi,$}\hspace{.3in}
\scalebox{.3}{\includegraphics{222ver.eps} } 
\raisebox{2.8ex}{= $-is\xi.$}
\end{center}

\begin{figure}[h]
\begin{center}
\scalebox{.4}{\includegraphics{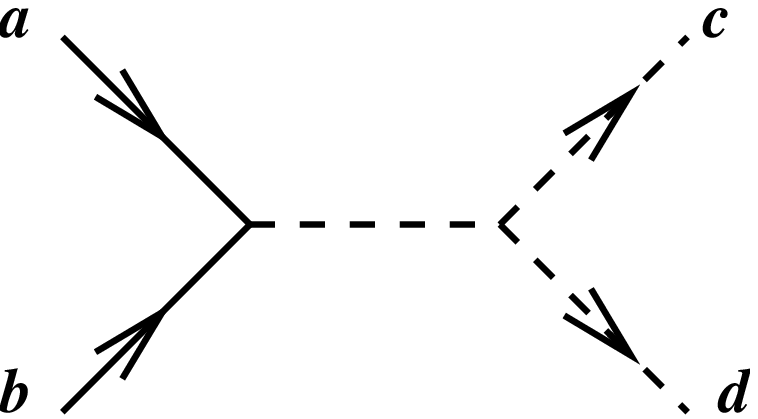}}
\raisebox{3.6ex}{
$\displaystyle =\frac{i\xi^2}{m^2}\frac{r}{(2-q+x)}$~,}\hspace {.2in}
\scalebox{.4}{\includegraphics{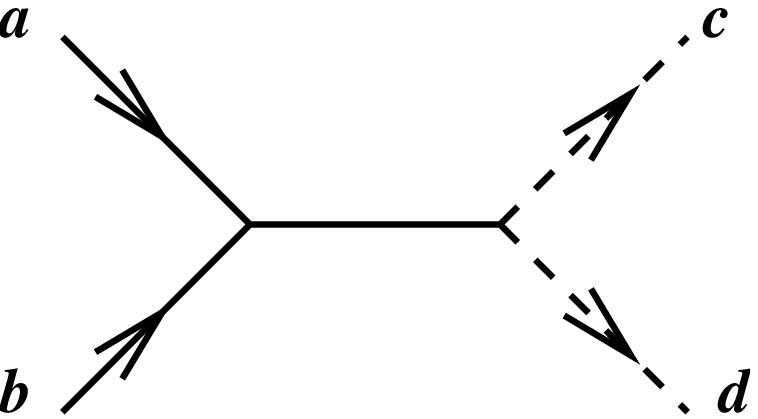}}
\raisebox{3.6ex}{$\displaystyle
  =-\frac{i\xi^2}{m^2}\frac{st}{(1+x)}$,~~~~~~~}
\caption{ a) and b) $2\phi_2\rightarrow 2\phi_1$ process}
\label{treecra}
\end{center}
\end{figure}

\begin{figure}[h]
\begin{center}
%%${}$\hspace {.2in}
\scalebox{.4}{\includegraphics{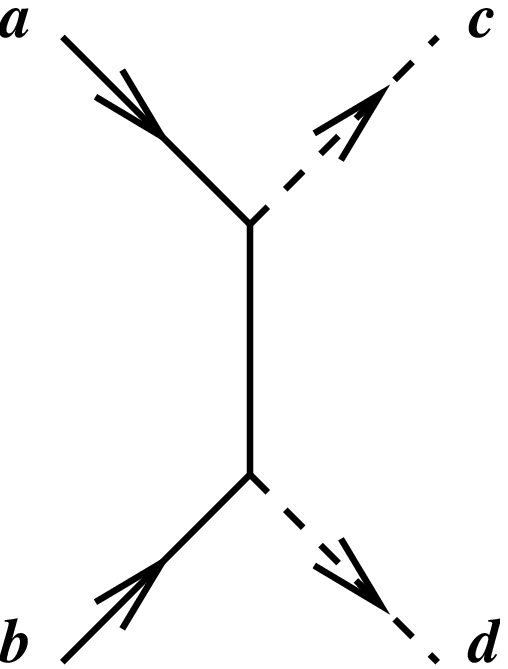}}\hspace{-.1in}
\raisebox{6.5ex}{
$\displaystyle =-\frac{i\xi^2}{m^2}\frac{1}
{(q-{\sqrt q}\,(ac^{-1}+a^{-1}c))}$~,}
\scalebox{.4}{\includegraphics{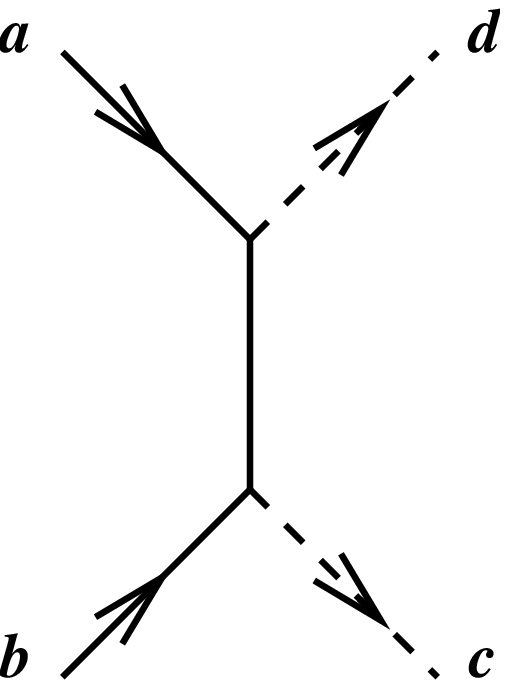}}\hspace{-0.1in}
\raisebox{6.5ex}{
$\displaystyle =-\frac{i\xi^2}{m^2}\frac{1}
{(q-{\sqrt q}\,(ad^{-1}+a^{-1}d))}$~.}
\end{center}\caption{a) and b) $2\phi_2\rightarrow 2\phi_1$ process
  with two $\phi_1\phi_2^2$ vertices}
\label{treecrb}
%%\end{figure}
%%\begin{figure}[h]
\begin{center}
%%${}$\hspace {.2in}
\scalebox{.4}{\includegraphics{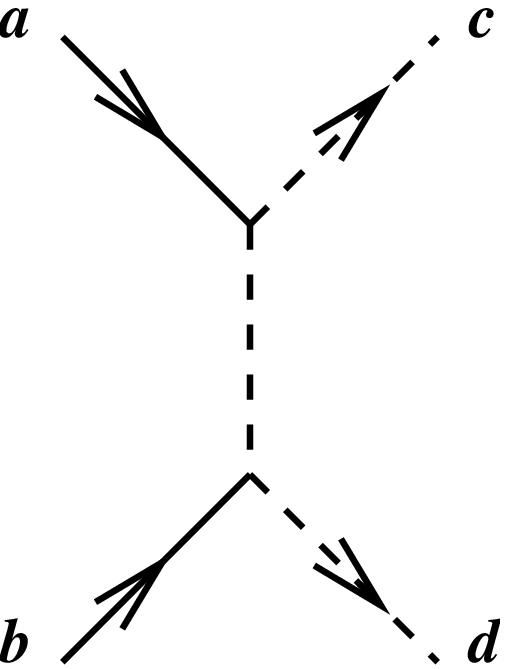}}\hspace{-.1in}
\raisebox{6.5ex}{
$\displaystyle =-\frac{i\xi^2}{m^2}\frac{t^2}
{(1-{\sqrt q}\,(ac^{-1}+a^{-1}c))}$~,}
\scalebox{.4}{\includegraphics{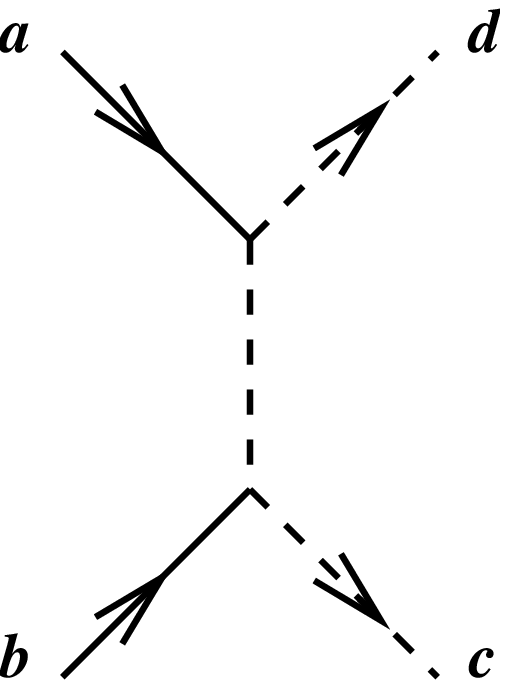}}\hspace{-0.1in}
\raisebox{6.5ex}{
$\displaystyle =-\frac{i\xi^2}{m^2}\frac{t^2}
{(1-{\sqrt q}\,(ad^{-1}+a^{-1}d))}$~.}
\end{center}\caption{a) and b) $2\phi_2\rightarrow 2\phi_1$ process
 with two $\phi_1^2\phi_2$ vertices}
\label{treecrc}
\end{figure}

We start with the process $2\phi_2\rightarrow 2\phi_1$, 
as done in the previous
section, calculate all possible tree level diagrams with the above Lagrangian.
For some particular combinations of $q,r,s$ and $t$ only the
contribution of the $2\phi_2\rightarrow 2\phi_1$ process comes
out to be a constant i.e. independent of $\emph{in}$ momenta,
and in that case this constant contribution
can be killed by adding a judiciously chosen
$\phi_1^2\phi_2^2$ term to the above Lagrangian (\ref{genl}).
This way one decides all possible three-point functions for
a particular theory to be constructed. Next one proceeds in manner
explained in the previous section, viz. studying the other tree
level processes and fixing the higher order interaction terms. 
Each of these combination of three-point functions (i.e. combination
of $q,r,s$ and $t$) gives an integrable model associated with a
rank 2 root system. In this section we would only fix the 3-point
couplings by studying $2\phi_2\rightarrow 2\phi_1$ process in detail. 
Following are the six diagrams (Figs. \ref{treecra}, \ref{treecrb},
\ref{treecrc}) contributing to the process 
$2\phi_2\rightarrow 2\phi_1.$ 

\noindent where $x\equiv ab^{-1}+a^{-1}b$. Using conservation of left- and
right-light-cone momenta, 
\begin{equation}
{\sqrt q}\,(c+d)=a+b~~~{\rm and}~~~
{\sqrt q}\,(c^{-1}+d^{-1})=a^{-1}+b^{-1}
\label{momconv}
\end{equation}
respectively one obtains
$x=q(cd^{-1}+c^{-1}d)+2(q-1)$.

\noindent Summing both the diagrams of Fig. \ref{treecrb}, we get
$\displaystyle \frac{i\xi^2}{m^2}\frac{(2-2q +x)}{(q^2-4q+2+x)}$,
using (\ref{momconv}).
Adding both the diagrams of Fig. \ref{treecrc}, we obtain
$\displaystyle \frac{i\xi^2}{m^2}\frac{t^2x}{(1-2q+qx)}$,
using (\ref{momconv}) again. Total contribution of all the six diagrams
then becomes,

\begin{equation}
\scalebox{.4}{\includegraphics{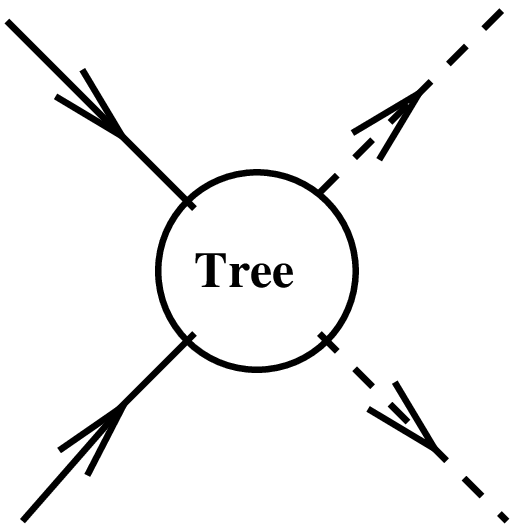}}\hspace{.1in}
\raisebox{5.5ex}{$\displaystyle=\frac{i\xi^2}{m^2}
\left[\frac{r}{(2-q+x)}-\frac{st}{(1+x)}
+\frac{(2-2q +x)}{(q^2-4q+2+x)}
+\frac{t^2x}{(1-2q+qx)}\right]$}
\label{2to2}
\end{equation}

Now one looks for cases for which expression (\ref{2to2}) is a
constant, i.e. independent of $x$ ( or incoming momenta ), so that 
it could be killed by
adding a $\phi_1^2\phi_2^2$ term to the Lagrangian
(\ref{genl}) with a suitably chosen coefficient.

{\bf Case I:} $t=0${\footnote{One need not fix $t=0$ ( or 
$r=s=0$ as done in the case II discussed later ) a priori. 
One could study the entire
expression (\ref{2to2}) as done in the case III later. The constancy 
constraint on (\ref{2to2}) would produce cases I and II as solutions.};

This gives a contribution (from the right hand side of (\ref{2to2})), 
$$\scalebox{.4}{\includegraphics{2to2.eps}}\hspace{.2in}
\raisebox{5.5ex}{$\displaystyle = \frac{i\xi^2}{m^2}\left[
\frac{r}{(2-q+x)}
+\frac{(2-2q +x)}{(q^2-4q+2+x)}\right],$}$$
\begin{equation}
= \frac{i\xi^2}{m^2}\left[
\frac{x^2 + x(4-3q+r) +(2q^2-6q+4+r(q^2-4q+2))}
{x^2 +x(q^2-5q+4)+(-q^3+6q^2-10q+4)}\right].
\label{case1}
\end{equation}
To have constant contribution one must now equate the coefficients of
various powers of $x$  in
numerator and denominator within the square bracket of expression 
(\ref{case1}).
In this case we get two equations. Equating coefficients of $x$ we get,
\begin{equation}
 q^2-2q=r,
\label{cefx}
\end{equation}
and equating the constants we have,
\begin{equation}
q^3 + q^2(r-4)+4q(1-r) +2r=0.
\label{cefc}
\end{equation} 
Using (\ref{cefx}) in (\ref{cefc}), we get three solutions, viz.
($q=0, r=0$); ($q=2, r=0$) and ($q=3, r=3$). 

{\bf a)} First of these, viz.($q=0, r=0$) is not acceptable as it sets mass of
particle $\phi_1$ as zero. 

{\bf b)} The second solution ($q=2, r=0$)
is already discussed in detail in the last section and leads to
{\boldmath $c_2^{(1)}$} ATFT. Moreover one has to choose $s=0$ for that.
It is clear that $s$ cannot be determined from the above 
as it vanishes from the expression once one chooses $t=0$ in 
(\ref{2to2}). Of course it can be fixed demanding zero contribution
from other tree level processes.
 
{\bf c)} The third solution ($q=3, r=3$) will lead to two different theories
depending on the values chosen for $s$. Allowed values of $s$ can be
again fixed by studying other tree level processes.

{\bf i)} If $s=0$, then the theory is {\boldmath $d_3^{(2)}$} with 
the Lagrangian,
\begin{equation}
{\cal L_{\mbox{\boldmath{$\scriptstyle d_3^{(2)}$}}}}={1\over 2}
\partial{\phibold}\cdot\partial{\phibold}-{m^2\over
\beta^2}\left[\sum_{i=0}^2e^{\beta{\albolds_i\cdot\,\phibolds}}-3\right],
\label{d32}
\end{equation} 
with simple and affine roots
$\albold_1=({\sqrt2},0),~\albold_2=(-\frac{1}
{\sqrt2},\frac{1}{\sqrt2}),
~\albold_0=(-\frac{1}
{\sqrt2},-\frac{1}
{\sqrt2})$ and $\beta={\sqrt2}\,\xi/m^2$.

{\bf ii)} If $s=\displaystyle -\frac{2}{\sqrt 3}$, then the theory is
{\boldmath $g_2^{(1)}$} and the corresponding Lagrangian becomes,
\begin{equation}
{\cal L}_{\mbox{\boldmath{$\scriptstyle g_2^{(1)}$}}}={1\over 2}
\partial{\phibold}\cdot\partial{\phibold}-{m^2\over
2\beta^2}\left[e^{\beta{\albolds_0\cdot\,\phibolds}}+
3e^{\beta{\albolds_1\cdot\,\phibolds}}+2
e^{\beta{\albolds_2\cdot\,\phibolds}}-6\right],
\label{g21}
\end{equation} 
with simple and affine roots
$\albold_1=(-\frac{1}{\sqrt2},\frac{1}{\sqrt6}),~\albold_2=({\sqrt2},0),
~\albold_0=(-\frac{1}{\sqrt2},-\frac{3}{\sqrt2})$ and 
$\beta={\sqrt2}\,\xi/m^2$.

{\bf Case II:} $r=s=0$;

In this case again we demand contribution ( from the right hand side
of (\ref{2to2})),  
$$\scalebox{.4}{\includegraphics{2to2.eps}}\hspace{.2in}
\raisebox{5.5ex}{$\displaystyle =\frac{i\xi^2}{m^2}\left[
\frac{(2-2q +x)}{(q^2-4q+2+x)}
+\frac{t^2x}{(1-2q+qx)}\right],$}$$
\begin{equation}
%%&=&\frac{(2-2q +x)}{(q^2-4q+2+x)}
%%+\frac{t^2x}{(1-2q+qx)}\\
=\frac{i\xi^2}{m^2}\left[
\frac{(1+{{t^2}\over q})x^2+(t^2q-4t^2-2q+{1\over q}+{2t^2\over
      q})x+(2-2q)({1\over q}-2)}{x^2+(q^2-4q+{1\over q})x+(q^2-4q+2)
({1\over q}-2)}\right],
\label{case2}
\end{equation}
for the process
$2\phi_2\rightarrow2\phi_1$ to be a constant.
Proceeding exactly like
the previous way (i.e. matching the coefficients of various powers 
$x$ in numerator and denominator) we have the following two distinct
solutions,
viz. ($\displaystyle q=\frac{3+{\sqrt 5}}{2} , t=\frac{1+{\sqrt
    5}}{2}$) and ($q=1,t^2=-1$).

{\bf a)} First solution, ($\displaystyle q=\frac{3+{\sqrt 5}}{2} , 
t=\frac{1+{\sqrt 5}}{2}$), leads to the 
{\boldmath $a_4^{(2)}$} ATFT. The Lagrangian for which is given by,

\begin{equation}
{\cal L}_{\mbox{\boldmath{$\scriptstyle a_4^{(2)}$}}}={1\over 2}
\partial{\phibold}\cdot\partial{\phibold}-{2m^2\over
5\beta^2}\frac{1}{(1-\sin2\theta)}\left[e^{\beta{\albolds_0\cdot\,\phibolds}}+
2e^{\beta{\albolds_1\cdot\,\phibolds}}+
2e^{\beta{\albolds_2\cdot\,\phibolds}}-5\right],
\label{a42}
\end{equation} 
where affine and simple roots are
$\albold_0=(-{\sqrt2}\sin(\pi/4+\theta),{\sqrt2}\cos(\pi/4+\theta))$,\\
$\albold_1=(\cos\theta,\sin\theta)$, $\albold_2=
(-\frac{1}{\sqrt2}\sin(\pi/4-\theta),-\frac{1}{\sqrt2}\cos(\pi/4-\theta))$,
with $2\tan 2\theta=1$ and $\displaystyle \beta=\frac{(1-\csc
  2\theta)}{(\sin\theta-\cos\theta)}\frac{\xi}{m^2}$.

{\bf b)} It is clear from expression (\ref{case2})
that the second solution ($q=1,t^2=-1$) will result in
vanishing contribution for the above process. This also asks for an
imaginary coupling $t$. This, we believe, should lead 
to {\boldmath $a^{(1)}_2$} ATFT
which is another rank 2 ATFT available with mass ratio of two fields
as unity (i.e. $q=1$). One must note that in {\boldmath $a^{(1)}_2$} 
theory the two fields are not self conjugate but mutually conjugate. 

{\bf Case III:} $r\neq 0, s\neq 0, t\neq 0$; 

In this case we obtain four equations (equating various powers of $x$
in the numerator and the denominator of the expression (\ref{2to2}),
as done earlier) by demanding the contribution
of the same $2\phi_2\rightarrow 2\phi_1$ process be a constant.
After some cumbersome algebra we reach the following solution,
($q=2+{\sqrt3},~ r=-3-2{\sqrt3},~ t=2+{\sqrt3}, ~s=-{\sqrt 3}$).
This leads to the last remaining rank 2 ATFT, viz. theory associated
with {\boldmath $d_4^{(3)}$} root system.
Lagrangian for which is,

\begin{equation}
{\cal L}_{\mbox{\boldmath{$\scriptstyle d_4^{(3)}$}}}={1\over 2}
\partial{\phibold}\cdot\partial{\phibold}-{{\sqrt3}m^2\over
2({\sqrt3}-1)\beta^2}\left[e^{\beta{\albolds_0\cdot\,\phibolds}}+
e^{\beta{\albolds_1\cdot\,\phibolds}}+
2e^{\beta{\albolds_2\cdot\,\phibolds}}-4\right],
\label{d43}
\end{equation} 
where simple and affine roots are
$\albold_1=(\frac{1+{\sqrt3}}{2},~\frac{1-{\sqrt3}}{2})$,
$\albold_2=(-\frac{1+{\sqrt3}}{2\sqrt3},~\frac{1-{\sqrt3}}{2\sqrt3})$,\\ 
$\albold_0=
(\frac{{\sqrt3}-1}{2\sqrt3},~\frac{{\sqrt3}+1}{2\sqrt3})$
and $\beta =-2{\sqrt 3}\,\xi/m^2$.

This completes our list of distinct solutions.
There are other solutions like
($\displaystyle q=\frac{3+{\sqrt 5}}{2} , t=-\frac{1+{\sqrt 5}}{2}$),
($\displaystyle q=\frac{3-{\sqrt 5}}{2} , t=\frac{1-{\sqrt 5}}{2}$)
and ($q=2-{\sqrt3},~ r=-3+2{\sqrt3},~ t=2-{\sqrt3}, ~s={\sqrt 3}$) etc.
which would keep expression (\ref{2to2}) constant but these 
are not distinct in a sense that they would not produce any new ATFT.
The first one could be obtained from the
case {\bf  II a)} by changing the field $\phi_2$ to $-\phi_2$.
The second of the above is again same as the solution {\bf  II a)}. In
this case the roles of the fields $\phi_1$ and $\phi_2$ are interchanged.
Both of these would lead to the same {\boldmath $a_4^{(2)}$} theory.
The last solution viz.($q=2-{\sqrt3},~ r=-3+2{\sqrt3},~ t=2-{\sqrt3},
~s={\sqrt 3}$) is again same as case {\bf III} with 
$\phi_1 \leftrightarrow \phi_2$ and leads to {\boldmath $d_4^{(3)}$} ATFT.

The above cases exhaust all possible solutions or 
acceptable values of $q,r,s$ and $t$ which will
respect elasticity and also exhaust
all possible ATFT associated with rank two root systems, viz.
{\boldmath $a_2^{(1)}$, $a_4^{(2)}$, $c_2^{(1)}$, 
$d_3^{(2)}$, $d_4^{(3)}$} and {\boldmath $g_2^{(1)}$}. $S$-matrices
and other details about these models can be found in Ref.[11-16]. 

Our Lagrangians for various ATFT may look little different from
the ones existing in the literature. This is due to the fact  
in the expressions of these Lagrangians we have chosen the 
simple and affine roots such a way that the mass matrix becomes
diagonal.

\section{Summary and Results}
\setcounter{equation}{0}

Here we summarize the results. In a pedagogical way we have introduced
the way of constructing integrable models in $1+1$ dimensions.
Starting with simple scalar field theories we have exploited the
elasticity property ( no particle production ) at tree level in 
the scattering of scalar particles for constructing affine Toda
field theory associated with rank one and rank two root systems.
It has been shown that the relative masses and three-point couplings 
could be fixed by vanishing amplitude of 
4-point function ( $2\phi_2\rightarrow 2\phi_1$ ) in case
of two scalar fields. 
We summarize the findings of the section 5 in the following table.

\begin{table}[h]
\caption{Relative strengths of the mass terms and the three-point
 couplings for rank 2 ATFT.}
\begin{center}
\begin{tabular}{l|c|c|c|c|c|c|c}\hline
 &\multicolumn{2}{|c|}{ mass terms}&
\multicolumn{4}{|c|}{Three-point interaction terms}&\\
\cline{2-7} 
{Case} & {$\phi_1^2$}& {$\phi_2^2$} & {$\phi_1^3$} & {$\phi_2^3$} &
{$\phi_1^2\phi_2$} & {$\phi_1\phi_2^2$} 
& {Theory}\\
\cline{2-7}
 & {$q$} & - & {$r$} & {$s$} & {$t$} &- & \\
\hline 
{\bf I b} & 2 & 1 & 0 & 0 & 0 & $-1$ & {\boldmath $c_2^{(1)}$}\\
{\bf I c i} & 3 & 1 & 3 & 0 & 0 & $-1$ & {\boldmath $d_3^{(2)}$}\\
{\bf I c ii} & 3 & 1 & 3 & $\displaystyle -\frac{2}{\sqrt 3}$ & 0 & $-1$ & 
{\boldmath $g_2^{(1)}$}\\
{\bf II a} & $\displaystyle \frac{3+{\sqrt 5}}{2}$ & 1 & 0 & 0 
&$\displaystyle \frac{1+{\sqrt 5}}{2}$  & $-1$ & {\boldmath $a_4^{(2)}$}\\
{\bf II b} & 1 & 1 & 0 & 0 & $\pm i$ & $-1$ & {\boldmath $a_2^{(1)}$}\\
{\bf III }& ${2+{\sqrt 3}}$ & 1 &  $-{3-2{\sqrt 3}}$ & $-{\sqrt 3}$ 
&${2+{\sqrt 3}}$ & $-1$ & {\boldmath $d_4^{(3)}$}\\ 
\hline
\end{tabular}
\end{center}
 \label{table3}
 \end{table}

Further it was shown
that once the three-point coupling are fixed, the higher
order couplings are determined uniquely by demanding vanishing of
various other scattering processes at tree level{\footnote {One can calculate
higher order couplings from three point couplings, see Ref. \cite{BCDSe}}}.
We have calculated 5-point functions and verified explicitly 
no particle production (i.e. vanishing amplitudes for 
2-$particle \rightarrow$ 3-$particle$ processes) in $a_2^{(2)}$
and $c_2^{(1)}$ ATFT for the very first time, we believe.
Each combination of allowed three-point couplings produces an ATFT
associated with a particular rank two root system. 
We strongly believe that the same procedure could also be used for
constructing ATFT  associated with root systems having rank greater
than two. It would be nice if one develops a way which works for
affine Toda field theories in general. Our effort is just a modest
beginning in this direction. 

{\bf Acknowledgements :} Author thanks Prof. Ryu Sasaki for reading
the manuscript and making valuable suggestions and comments at various
stages of this work. Author also thanks Mr. Tapan Naskar who tried
evaluating some of the Feynman diagrams at the earlier stage of this work.

\section*{Appendix A1:}
\setcounter{equation}{0}
\renewcommand{\theequation}{A1.\arabic{equation}}
\begin{eqnarray}
&\raisebox{5.5ex}{1)}&\hspace{.4in}
\scalebox{.40}{\includegraphics{tree331.eps}}\hspace{.3in}
\raisebox{5.5ex}{$\displaystyle =-\frac{i\eta^3}{6m^4}
\frac{1}{(2-c-c^{-1})}$}
\label{appena1}\\
&\raisebox{3.5ex}{2)}&\hspace{.4in}
\scalebox{.40}{\includegraphics{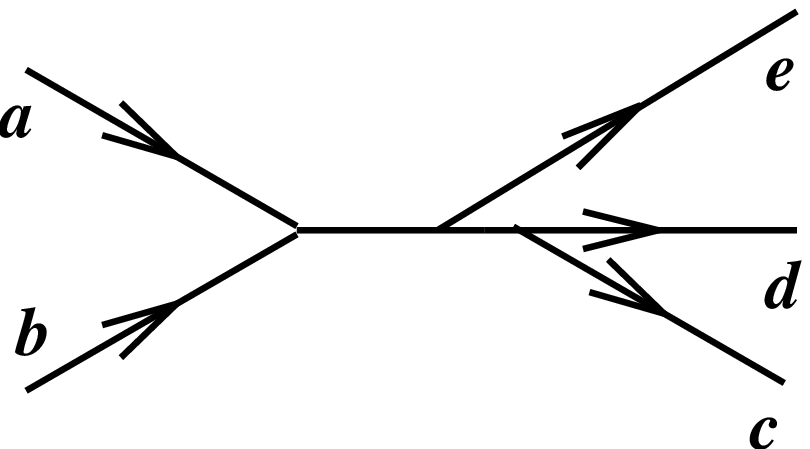}}\hspace{.3in}
\raisebox{5.0ex}{$\displaystyle =\frac{i\eta^3}{6m^4}
\frac{(1+\delta)}{\delta^2}$}
\label{appena2}\\
&\raisebox{3.5ex}{3)}&\hspace{.4in}
\scalebox{.45}{\includegraphics{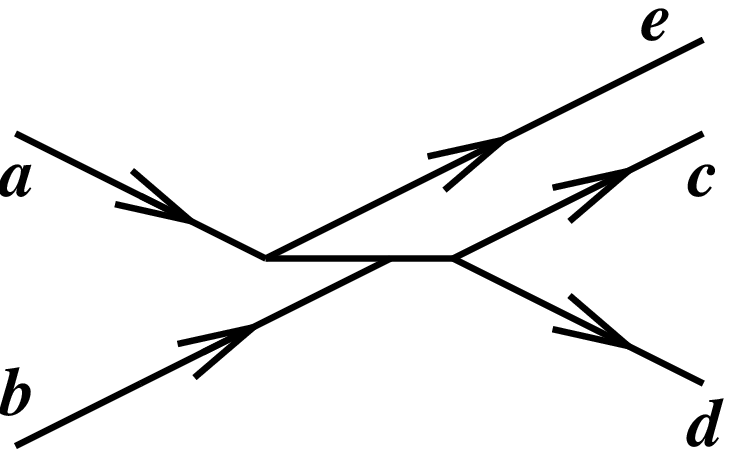}}\hspace{.3in}
\raisebox{5.5ex}{$\displaystyle =-\frac{i\eta^3}{2m^4}
\frac{(1+\delta)^2}{\delta^2(1+\delta+\delta^2)}$}
\label{appena3}\\
&\raisebox{3.5ex}{4)}&\hspace{.4in}
\scalebox{.45}{\includegraphics{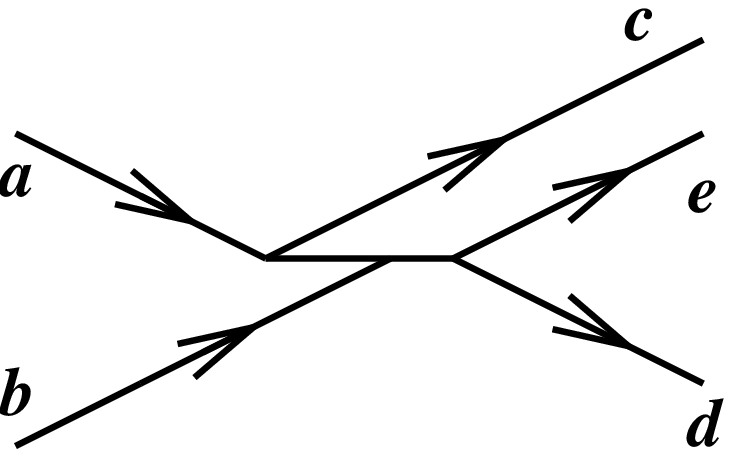}}\hspace{.3in}
\raisebox{4.5ex}{$\displaystyle =-\frac{i\eta^3}{2m^4}
\frac{1}{(1-c-c^{-1})(2-c-c^{-1})}$}
\label{appena4}\\
&\raisebox{3.5ex}{5)}&\hspace{.4in}
\scalebox{.45}{\includegraphics{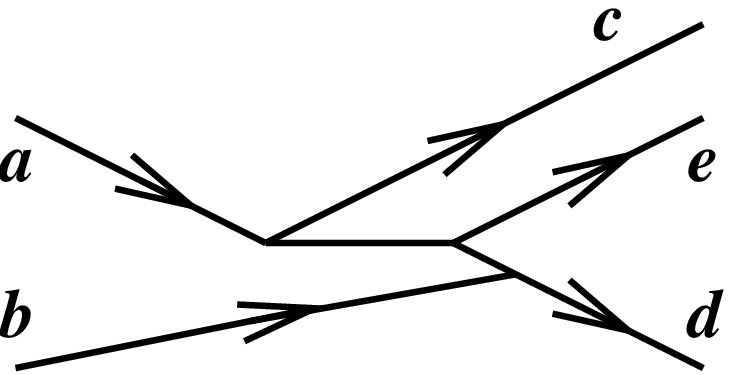}}\hspace{.3in}
\raisebox{3.5ex}{$\displaystyle =-\frac{i\eta^3}{m^4}
\frac{1}{(1-c-c^{-1})(1-d-d^{-1})}$}
\label{appena5}\\
&\raisebox{3.5ex}{6)}&\hspace{.4in}
\scalebox{.45}{\includegraphics{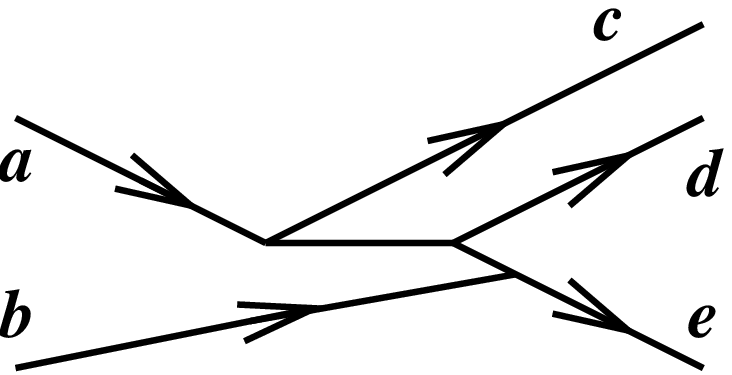}}\hspace{.3in}
\raisebox{3.5ex}{$\displaystyle =\frac{i\eta^3}{m^4}
\frac{1}{(1-c-c^{-1})}\frac{(1+\delta)}{(1+\delta+\delta^2)}$}
\label{appena6}\\
&{7)}&\hspace{.4in} = 4)~~(c\longleftrightarrow d) 
=-\frac{i\eta^3}{2m^4}
\frac{1}{(1-d-d^{-1})(2-d-d^{-1})}\\
&{8)}&\hspace{.4in} = 5)~~(c\longleftrightarrow d) 
=-\frac{i\eta^3}{m^4}\frac{1}{(1-c-c^{-1})(1-d-d^{-1})}\\
&{9)}&\hspace{.4in} = 6)~~(c\longleftrightarrow d) 
=\frac{i\eta^3}{m^4}
\frac{1}{(1-d-d^{-1})}
\frac{(1+\delta)}{(1+\delta+\delta^2)}\\
&{10)}&\hspace{.4in} = 1)~~(c\longleftrightarrow d) 
=-\frac{i\eta^3}{6m^4}\frac{1}{(2-d-d^{-1})}\\
&{11)}&\hspace{.4in} = 6)~~(a\longleftrightarrow b) 
=\frac{i\eta^3}{m^4}\frac{1}{(1-c-c^{-1})}
\frac{(1+\delta)}{(1+\delta+\delta^2)}\\
&{12)}&\hspace{.4in} = 9)~~(a\longleftrightarrow b) 
=\frac{i\eta^3}{m^4}\frac{1}{(1-d-d^{-1})}
\frac{(1+\delta)}{(1+\delta+\delta^2)}\\
&{13)}&\hspace{.4in} = 3)~~(a\longleftrightarrow b) 
=-\frac{i\eta^3}{2m^4}\frac{(1+\delta)^2}{\delta^2(1+\delta+\delta^2)}\\
&{14)}&\hspace{.4in} = 4)~~(a\longleftrightarrow b) 
=-\frac{i\eta^3}{2m^4}\frac{1}{(1-c-c^{-1})(2-c-c^{-1})}\\
&{15)}&\hspace{.4in} = 14)~~(c\longleftrightarrow d) 
=-\frac{i\eta^3}{2m^4}\frac{1}{(1-d-d^{-1})(2-d-d^{-1})}
\end{eqnarray}

\section*{Appendix A2:}
\setcounter{equation}{0}
\renewcommand{\theequation}{A2.\arabic{equation}}
\begin{eqnarray}
&\raisebox{5.5ex}{1)}&\hspace{.4in}
\scalebox{.5}{\includegraphics{tree34.eps}}\hspace{.3in}
\raisebox{5.5ex}{$\displaystyle =-\frac{i\eta\lambda}{3m^2}$}
\label{appenb1}\\
&\raisebox{3.5ex}{2)}&\hspace{.4in}
\scalebox{.55}{\includegraphics{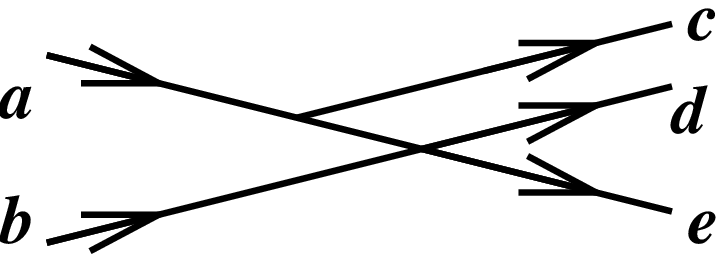}}\hspace{.3in}
\raisebox{3.5ex}{$\displaystyle =-\frac{i\eta\lambda}{m^2}
\frac{1}{(1-c-c^{-1})}$}
\label{appenb2}\\
&\raisebox{3.5ex}{3)}&\hspace{.4in}
\scalebox{.55}{\includegraphics{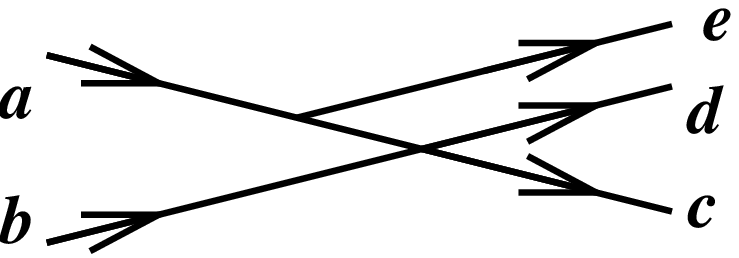}}\hspace{.3in}
\raisebox{3.5ex}{$\displaystyle =\frac{i\eta\lambda}{m^2}
\frac{(1+\delta)}{(1+\delta+\delta^2)}$}
\label{appenb3}\\
&\raisebox{3.5ex}{4)}&\hspace{.4in}
\scalebox{.55}{\includegraphics{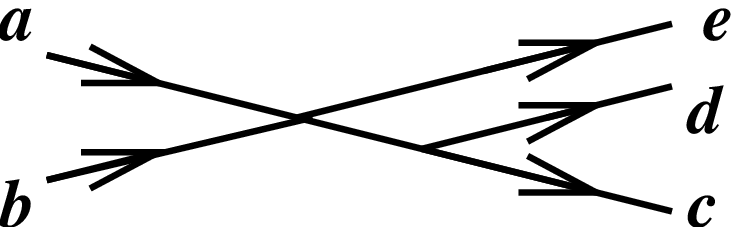}}\hspace{.3in}
\raisebox{3.5ex}{$\displaystyle =\frac{i\eta\lambda}{2m^2}
\frac{(1+\delta)}{\delta^2}$}
\label{appenb4}\\
&\raisebox{3.5ex}{5)}&\hspace{.4in}
\scalebox{.55}{\includegraphics{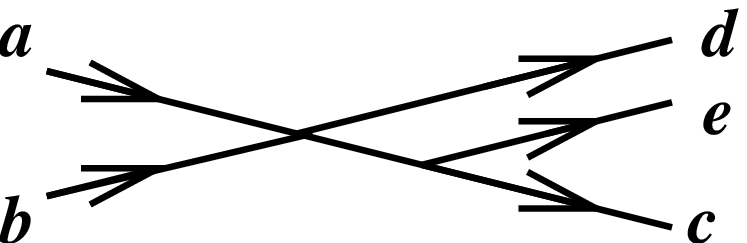}}\hspace{.3in}
\raisebox{3.5ex}{$\displaystyle =-\frac{i\eta\lambda}{2m^2}
\frac{1}{(2-d-d^{-1})}$}
\label{appenb5}\\
&{6)}&\hspace{.4in} = 2)~~(a\longleftrightarrow b) 
=-\frac{i\eta\lambda}{m^2}\frac{1}{(1-c-c^{-1})}\\
&{7)}&\hspace{.4in} = 2)~~(c\longleftrightarrow d) 
=-\frac{i\eta\lambda}{m^2}\frac{1}{(1-d-d^{-1})}\\
&{8)}&\hspace{.4in} = 7)~~(a\longleftrightarrow b) 
=-\frac{i\eta\lambda}{m^2}\frac{1}{(1-d-d^{-1})}\\
&{9)}&\hspace{.4in} = 3)~~(a\longleftrightarrow b) 
=\frac{i\eta\lambda}{m^2}
\frac{(1+\delta)}{(1+\delta+\delta^2)}\\
&{10)}&\hspace{.4in} = 5)~~(c\longleftrightarrow d) 
=-\frac{i\eta\lambda}{2m^2}
\frac{1}{(2-c-c^{-1})}
\end{eqnarray}

\section*{Appendix B1:}
\setcounter{equation}{0}
\renewcommand{\theequation}{B1.\arabic{equation}}
\begin{eqnarray}
&\raisebox{5.5ex}{1)}&\hspace{.4in}
\scalebox{.55}{\includegraphics{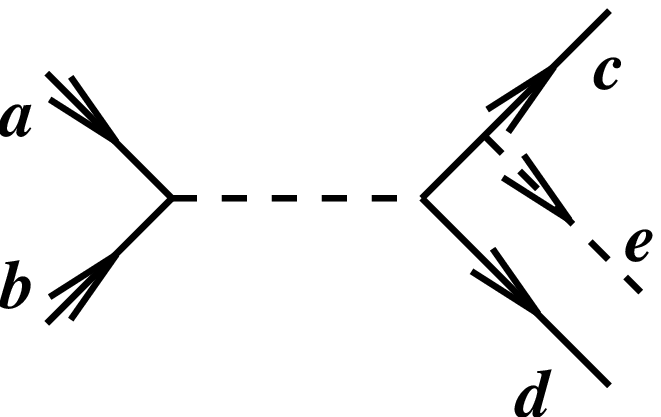}}\hspace{.3in}
\raisebox{6.5ex}{$\displaystyle =\frac{i\xi^3}{4m^4}
\frac{1}{(2-d-d^{-1})}$}
\label{appenc1}\\
&\raisebox{3.5ex}{2)}&\hspace{.4in}
\scalebox{.55}{\includegraphics{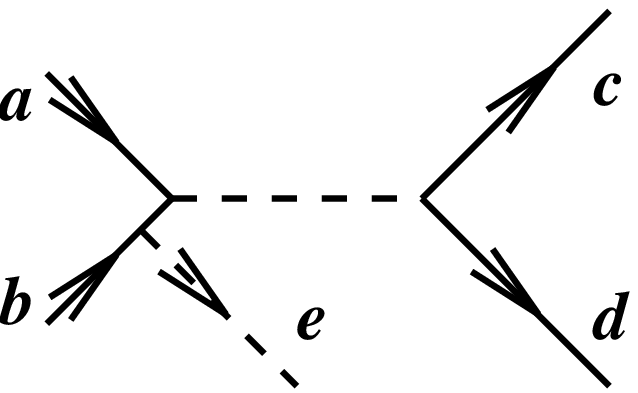}}\hspace{.3in}
\raisebox{4.5ex}{$\displaystyle =\frac{i\xi^3}{4m^4}
\frac{e^2}{(1-{\sqrt 2}\,e+e^2)^2}$}
\label{appenc2}\\
&{3)}&\hspace{.4in} = 1)~~(c\longleftrightarrow d) 
=\frac{i\xi^3}{4m^4}
\frac{1}{(2-c-c^{-1})}\\
&{4)}&\hspace{.4in} = 2)~~(a\longleftrightarrow b) 
=\frac{i\xi^3}{4m^4}
\frac{e^2}{(1-{\sqrt 2}\,e+e^2)^2}\\
&\raisebox{8.5ex}{5)}&\hspace{.4in}
\scalebox{.5}{\includegraphics{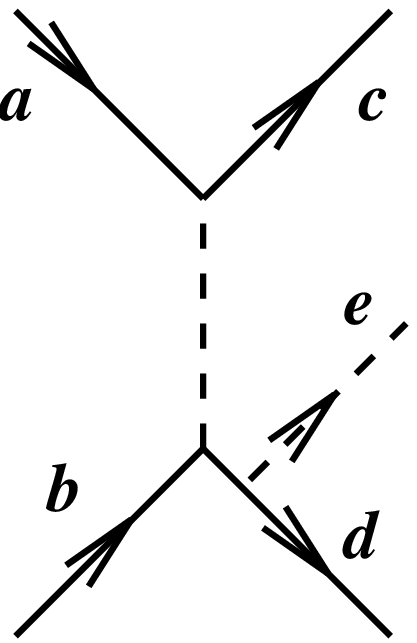}}\hspace{.3in}
\raisebox{8.5ex}{$\displaystyle =-\frac{i\xi^3}{2m^4}
\frac{1}{(c+c^{-1})(2-c-c^{-1})}$}
\label{appenc5}\\
&\raisebox{8.5ex}{6)}&\hspace{.4in}
\scalebox{.5}{\includegraphics{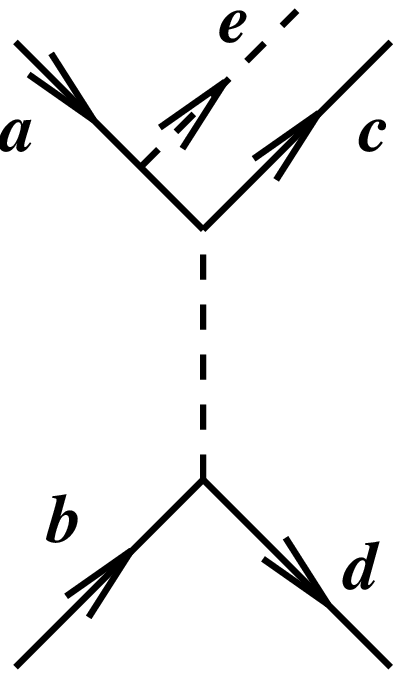}}\hspace{.3in}
\raisebox{8.5ex}{$\displaystyle =\frac{i\xi^3}{{\sqrt2}\,m^4}
\frac{1}{(d+d^{-1})(e+e^{-1}-{\sqrt2})}$}
\label{appenc6}\\
&{7)}&\hspace{.4in} = 5)~~(a\longleftrightarrow b) 
=-\frac{i\xi^3}{2m^4}\frac{1}{(c+c^{-1})(2-c-c^{-1})}\\
&{8)}&\hspace{.4in} = 5)~~(c\longleftrightarrow d) 
=-\frac{i\xi^3}{2m^4}\frac{1}{(d+d^{-1})(2-d-d^{-1})}\\
&{9)}&\hspace{.4in} = 7)~~(c\longleftrightarrow d) 
=-\frac{i\xi^3}{2m^4}\frac{1}{(d+d^{-1})(2-d-d^{-1})}\\
&{10)}&\hspace{.4in} = 6)~~(a\longleftrightarrow b) 
=\frac{i\xi^3}{{\sqrt2}\,m^4}
\frac{1}{(d+d^{-1})(e+e^{-1}-{\sqrt2})}\\
&{11)}&\hspace{.4in} = 6)~~(c\longleftrightarrow d) 
=\frac{i\xi^3}{{\sqrt2}\,m^4}
\frac{1}{(c+c^{-1})(e+e^{-1}-{\sqrt2})}\\
&{12)}&\hspace{.4in}= 10)~~(c\longleftrightarrow d) =
\frac{i\xi^3}{{\sqrt2}\,m^4}
\frac{1}{(c+c^{-1})(e+e^{-1}-{\sqrt2})}\\
&\raisebox{5.5ex}{13)}&\hspace{.4in}
\scalebox{.55}{\includegraphics{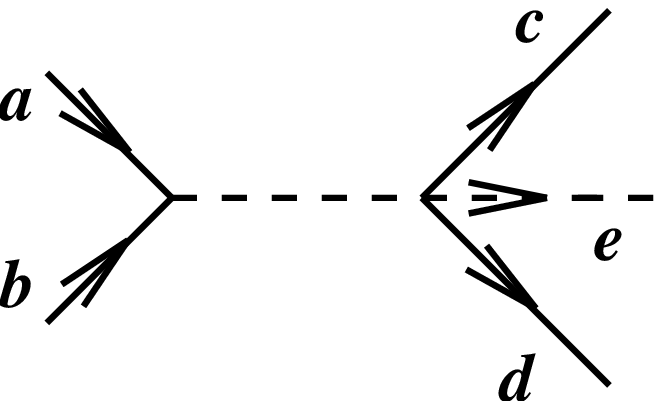}}\hspace{.3in}
\raisebox{6.0ex}{$\displaystyle =\frac{i\xi^3}{2m^4}$}
\label{append1}\\
&\raisebox{5.5ex}{14)}&\hspace{.4in}
\scalebox{.55}{\includegraphics{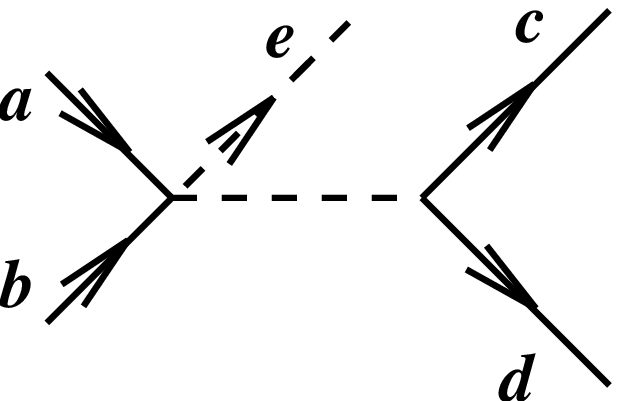}}\hspace{.3in}
\raisebox{5.5ex}{$\displaystyle =-\frac{i\xi^3}{2{\sqrt2}m^4}
\frac{1}{(e+e^{-1}-{\sqrt 2})}$}
\label{append2}\\
&\raisebox{8.5ex}{15)}&\hspace{.4in}
\scalebox{.55}{\includegraphics{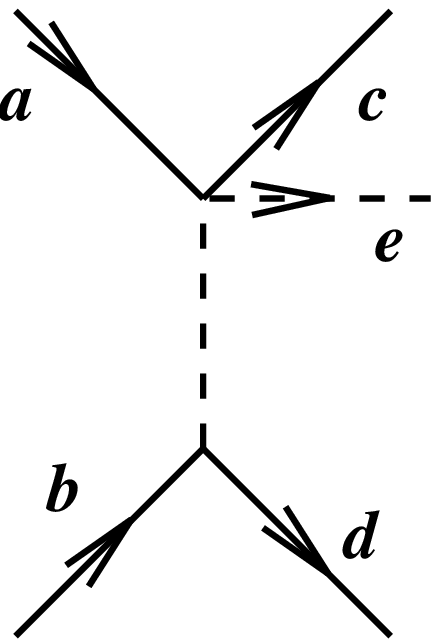}}\hspace{.3in}
\raisebox{8.5ex}{$\displaystyle =-\frac{i\xi^3}{m^4}
\frac{1}{(d+d^{-1})}$}
\label{append3}\\
&{16)}&\hspace{.4in} = 15)~~(a\longleftrightarrow b) 
=-\frac{i\xi^3}{m^4}\frac{1}{(d+d^{-1})}\\
&{17)}&\hspace{.4in} = 15)~~(c\longleftrightarrow d) 
=-\frac{i\xi^3}{m^4}\frac{1}{(c+c^{-1})}\\
&{18)}&\hspace{.4in} = 16)~~(c\longleftrightarrow d) 
=-\frac{i\xi^3}{m^4}\frac{1}{(c+c^{-1})}
\end{eqnarray}

\section*{Appendix B2:}
\setcounter{equation}{0}
\renewcommand{\theequation}{B2.\arabic{equation}}
\begin{eqnarray}
&\raisebox{5.4ex}{1)}&\hspace{.4in}
\scalebox{.55}{\includegraphics{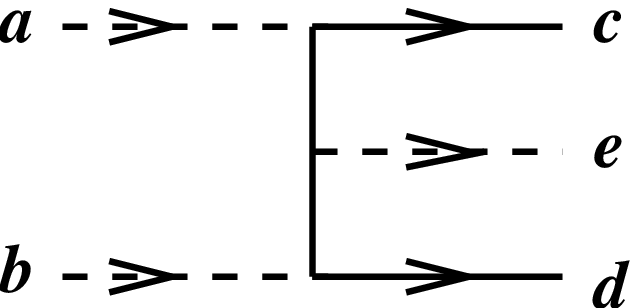}}\hspace{.3in}
\raisebox{6.5ex}{$\displaystyle =\frac{i\xi^3}{m^4}
\frac{1}{(2-{\sqrt2}(c+c^{-1}))}\frac{1}{(2-{\sqrt2}(d+d^{-1}))}$}
\label{appene1}\\
&\raisebox{5.5ex}{2)}&\hspace{.4in}
\scalebox{.55}{\includegraphics{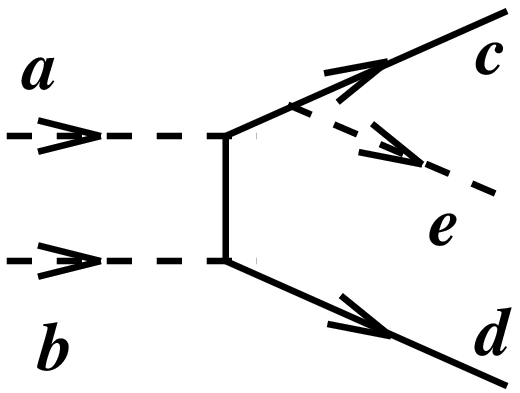}}\hspace{.3in}
\raisebox{4.5ex}{$\displaystyle =-\frac{i\xi^3}{4m^4}
\frac{1}{({\sqrt2}-d-d^{-1})}\frac{1}{(2{\sqrt2}-d-d^{-1})}$}
\label{appene2}\\
&{3)}&\hspace{.4in} = 1)~~(c\longleftrightarrow d) 
=\frac{i\xi^3}{m^4}
\frac{1}{(2-{\sqrt2}(c+c^{-1}))}\frac{1}{(2-{\sqrt2}(d+d^{-1}))}\\
&{4)}&\hspace{.4in} = 2)~~(a\longleftrightarrow b) 
=-\frac{i\xi^3}{4m^4}
\frac{1}{({\sqrt2}-d-d^{-1})}\frac{1}{(2{\sqrt2}-d-d^{-1})}\\
&{5)}&\hspace{.4in} = 2)~~(c\longleftrightarrow d) 
=-\frac{i\xi^3}{4m^4}
\frac{1}{({\sqrt2}-c-c^{-1})}\frac{1}{(2{\sqrt2}-c-c^{-1})}\\
&{6)}&\hspace{.4in} = 5)~~(a\longleftrightarrow b) 
=-\frac{i\xi^3}{4m^4}
\frac{1}{({\sqrt2}-c-c^{-1})}\frac{1}{(2{\sqrt2}-c-c^{-1})}\\
&\raisebox{5.5ex}{7)}&\hspace{.4in}
\scalebox{.5}{\includegraphics{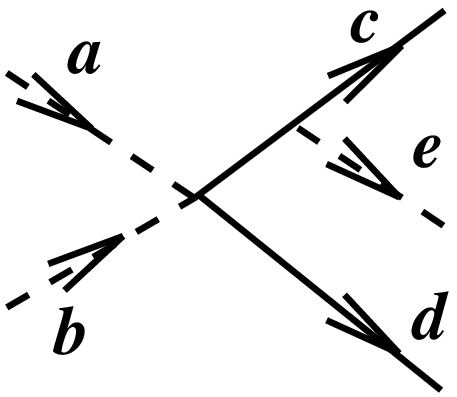}}\hspace{.3in}
\raisebox{5.5ex}{$\displaystyle =\frac{i\xi^3}{2{\sqrt2}\,m^4}
\frac{1}{(2{\sqrt2}-d-d^{-1})}$}
\label{appene3}\\
&\raisebox{5.5ex}{8)}&\hspace{.4in}
\scalebox{.5}{\includegraphics{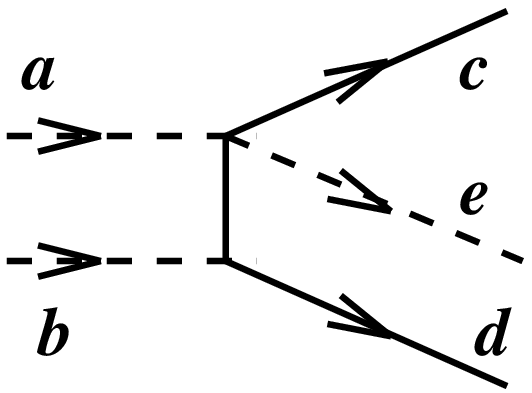}}\hspace{.3in}
\raisebox{5.5ex}{$\displaystyle =\frac{i\xi^3}{{\sqrt2}\,m^4}
\frac{1}{({\sqrt2}-d-d^{-1})}$}
\label{appene4}\\
&{9)}&\hspace{.4in} = 7)~~(c\longleftrightarrow d) 
=\frac{i\xi^3}{2{\sqrt2}\,m^4}
\frac{1}{(2{\sqrt2}-c-c^{-1})}\\
&{10)}&\hspace{.4in} = 8)~~(c\longleftrightarrow d) 
=\frac{i\xi^3}{{\sqrt2}\,m^4}
\frac{1}{({\sqrt2}-c-c^{-1})}\\
&{11)}&\hspace{.4in} = 8)~~(a\longleftrightarrow b) 
=\frac{i\xi^3}{{\sqrt2}\,m^4}
\frac{1}{({\sqrt2}-d-d^{-1})}\\
&{12)}&\hspace{.4in} = 11)~~(c\longleftrightarrow d) 
=\frac{i\xi^3}{{\sqrt2}\,m^4}
\frac{1}{({\sqrt2}-c-c^{-1})}
\end{eqnarray}
%%%%%%%%%%%%%%%%%%%%%%%%%%%%%%%%%%%%%%%%%%%%%%%%%%%%%%

\end{document}